\documentclass[aps,prd,twocolumn,showpacs,floatfix,preprintnumbers,nofootinbib,superscriptaddress]{revtex4}

\usepackage{bm}
\usepackage{amsfonts}
\usepackage{amsmath}
\usepackage{graphicx}
\usepackage[amssymb]{SIunits}
\usepackage{hyperref}

\newcommand{\eref}[1]{equation~(\ref{#1})}
\newcommand{\fref}[1]{Figure~\ref{#1}}
\newcommand{\sref}[1]{Section~\ref{#1}}
\newcommand{\tref}[1]{Table~\ref{#1}}
\newcommand{\NH}{\textrm{NH}}
\newcommand{\IH}{\textrm{IH}}
\renewcommand{\S}{\mathbf{S}}
\newcommand{\D}{\mathbf{D}}
\renewcommand{\P}{\mathbf{P}}
\newcommand{\B}{\mathbf{B}}
\newcommand{\M}{\mathbf{M}}
\newcommand{\g}{\mbox{\boldmath$\gamma$}}

\begin{document}


\title{Chaotic flavor evolution in an interacting neutrino gas}

\author{Rasmus Sloth Hansen}
\affiliation{Department of Physics and Astronomy,
 University of Aarhus, 8000 Aarhus C, Denmark}
\affiliation{School of Physics, The University of New South Wales, 
Sydney NSW 2052, Australia}
\author{Steen Hannestad}
\affiliation{Department of Physics and Astronomy,
 University of Aarhus, 8000 Aarhus C, Denmark}
\affiliation{Aarhus Institute of Advanced Studies,
 University of Aarhus, 8000 Aarhus C, Denmark}

\date{\today}

\begin{abstract}

Neutrino-neutrino refraction can lead to non-periodic flavor oscillations in dense neutrino gases, and it has been hypothesized that some solutions are chaotic in nature.
This is of particular interest in the case of neutrino emission from core-collapse supernovae where the measurement of the spectral shape for different flavors can provide crucial information about both neutrino physics and the physical conditions close to the proto-neutron star.
Whether a system is chaotic or not can be assessed by the Lyapunov exponents which quantify the rate of divergence of nearby trajectories in the system. We have done a numerical case study for a simple toy model of two neutrino flavors with two momentum states traveling against each other which is known to exhibit flavor transition instabilities. We find the leading Lyapunov exponent to be positive in all cases, confirming the chaoticity of the system for both the normal and the inverted neutrino mass hierarchy.
However, more Lyapunov exponents were approximately zero in the inverted hierarchy compared to the normal which has implications for the stability of the system.
To investigate this, we have calculated a generalized set of normal modes, the so-called covariant Lyapunov vectors. The covariant Lyapunov vectors associated with vanishing Lyapunov exponents showed the existence of marginally stable directions in phase space for some cases. 
While our analysis was done for a toy model example, it should work equally well for more realistic cases of neutrinos streaming from a proto-neutron star and provide valuable insight into the nature of the flavor instability.
We finally stress that our approach captures many more properties of the physical system than the linear stability analyses which have previously been performed.

\end{abstract}

\pacs{14.60.Pq, 97.60.Bw, 05.45.Pq}

\maketitle

\section{Introduction}                        
\label{sec:introduction}

It has been known for many years that the presence of a neutrino background can lead to highly non-trivial behavior of neutrino oscillations~\cite{Pantaleone1992, Samuel:1993uw}. The effect was first studied in the context of neutrino oscillations in the early Universe~\cite{Stodolsky:1986dx}, but it also manifests itself in the context of core-collapse supernovae~(see e.g.\ \cite{Hannestad:2006nj, Duan:2005cp, Duan:2006an, EstebanPretel:2007ec, Fogli:2007bk}). 
The origin of this phenomenon is the neutrino-neutrino interaction, which results in a background potential for the propagation of the neutrinos themselves. Contributions to the background potential come with a factor of $(1-\vec{v}_i\vec{v}_j)$, where $\vec{v}_i$ is the direction of the propagating neutrino momentum and $\vec{v}_j$ is the direction of the background neutrino momentum.
In a highly isotropic environment, such as the early Universe, one would expect that this term can be integrated out giving a tremendous simplification of the computational problem, and even for an only approximately spherical core-collapse supernova 
it seems reasonable to assume axial symmetry around a radial ray of neutrinos and integrate out the corresponding angles~\cite{Duan:2006an}.
The latter however, fails to capture an important class of solutions. This was first realized through a linearized stability analysis by Raffelt, Sarikas and Seixas~\cite{Raffelt:2013rqa} where they found a new flavor instability for the normal neutrino mass hierarchy. Their results have later been confirmed in various numerical simulations~\cite{Mirizzi:2013rla, Mirizzi:2013wda, Chakraborty:2014nma}, but important conceptual advances have also been made.

Shortly after the first instability had been discovered, it was shown that a similar instability can be found in a much simpler system \cite{Raffelt:2013isa}. Considering only two momentum states traveling in opposite directions, these authors showed the well known flavor inverting bipolar oscillations to be present in the inverted hierarchy, if the two states were prepared identically, while the flavor inversion was found in the normal hierarchy if the two states were prepared in a certain anti-symmetric way. For both mass hierarchies they found that even small perturbations to an otherwise stable solution would excite the bipolar oscillations and shortly thereafter give rise to a seemingly chaotic behavior.
Similar systems were analyzed by Sawyer~\cite{Sawyer:2008zs} from a somewhat different perspective, and he also found highly non-periodic solutions.

If one wants to address the question of chaoticity and go beyond a stationary stability analysis, it is not possible to do a full analytical analysis. Instead by using a numerical scheme, it is possible to calculate a variety of characteristic quantities. 
The Lyapunov exponent is such a characteristic quantity, generalizing the concept of exponential growth rates from the stationary stability analysis to any non-stationary solution. The Lyapunov exponents describe how small perturbations to a given solution will grow and thereby indicate if a system is chaotic or not~\cite{Hannestad:2013pha, Braad}. The associated covariant Lyapunov vectors generalize the concept of normal modes to periodic and even non-periodic trajectories, and from them it is possible to get information about which directions are expanding and contracting in phase space.

In this paper, we will calculate the Lyapunov exponents and covariant Lyapunov vectors for the two beam toy model for neutrino flavor oscillations with two opposite momentum states and discuss their implications.
First we will give an introduction to the two beam model in \sref{sec:model}. Then we will give a short description of Lyapunov analysis in \sref{sec:lyalight} referring the interested reader to the appendices and references for a more in depth discussion of the subject. In \sref{sec:num} we will present and discuss the results for the stationary-, bipolar-, and non-periodic cases. Finally we will have a few concluding remarks in \sref{sec:conclusions}.



\section{The two beam model}
\label{sec:model}

Our model contains only two momentum states, but before we specialize to that case, we will consider the more general case of $N$ momentum modes.

\subsection{N momentum modes}

In the general case we consider a neutrino gas of oscillating $\nu_e$ and $\nu_x$ ($x = \mu$ or $\tau$) consisting of $N$ momentum modes. Using the polarization vector parameterization of the density matrices, the oscillation equations without a matter background can be written as \cite{Raffelt:2013isa}
\begin{equation}
\label{eq:gendiff}
\begin{aligned}
  \dot{\mathbf{P}}_i &= \left( \omega_i \mathbf{B} + \frac{\mu}{2} \sum_{j=1}^N \left(\mathbf{P}_j - \bar{\mathbf{P}}_j\right) (1 - \vec{v}_j \cdot \vec{v}_i) \right) \times \mathbf{P}_i ,\\
  \dot{\bar{\mathbf{P}}}_i &= \left( -\omega_i \mathbf{B} + \frac{\mu}{2} \sum_{j=1}^N \left(\mathbf{P}_j - \bar{\mathbf{P}}_j\right) (1 - \vec{v}_j \cdot \vec{v}_i) \right) \times \bar{\mathbf{P}}_i ,
\end{aligned}
\end{equation}
where $\bar\P_i$ refers to antineutrinos, $\mu \sim 2\sqrt{2} G_F n_\nu$\footnote{Note that we define $\mu$ slightly different than in \cite{Raffelt:2013isa} to absorb a factor of 2 in \eref{eq:diff}.}, $\omega_i = \Delta m^2/2E_i$ , $\mathbf{B}$ is the mass unit vector in flavor space, and $\vec{v}_i = \vec{p}_i/E_i$ is the direction of the momentum. We use arrows to indicate vectors in real space while bold faces refer to vectors in polarization space.

There are two obvious choices for the coordinate system in polarization space. The first takes the $z$-direction to coincide with the pure electron neutrino state, and the second lets $\mathbf{B}$ determine the orientation of the $z$-axis and exploits the symmetries of the equations. Since we do not aim to calculate any oscillation probabilities, we will adopt the latter convention and set $\mathbf{B} = (0,0,-1)$. This choice for $\B$ ensures that $\omega > 0$ corresponds to the normal hierarchy while $\omega < 0$ corresponds to the inverted hierarchy.

Since the two-flavor oscillation is a two level system, it has many similarities to spins, and this is highlighted by the formulation in terms of polarization vectors.
In the isospin convention, the isospin vectors can be identified with angular momenta, but this means that neutrinos and antineutrinos with similar flavor content will be associated with isospin vectors pointing in opposite directions. To avoid this, we choose the opposite sign for $\bar\P$, and therefore $\P_i$ and $-\bar\P_i$ correspond to the angular momenta.

As we assume no dissipation in our equations of motion, the system is Hamiltonian, and it turns out to be enlightening to consider a classical Hamiltonian formulation of the equations.
For this kind of motion confined to a set of spheres (the lengths of $\P_i$ and $\bar\P_i$ are constant), the canonical coordinates and momenta are $\phi_i$ and $P_{iz} = P_i\cos\theta_i$ for the neutrinos and $\bar{\phi}_i$ and $-\bar{P}_{iz} = \bar{P}_i \cos\bar{\theta}_i$ for the antineutrinos. From these variables we can define the polarization vectors as 
\begin{equation}
\begin{aligned}
\P_i &= P_i(\cos\phi_i\sin\theta_i, \sin\phi_i\sin\theta_i, \cos\theta_i),\\
\bar{\P}_i &= -\bar P_i (\cos\bar\phi_i\sin\bar\theta_i, \sin\bar\phi_i\sin\bar\theta_i, \cos\bar\theta_i).
\end{aligned}
\end{equation}
Furthermore, we can derive the Poisson brackets $\left\{P_{ia}, P_{ib}\right\} = \epsilon_{abc}P_{ic}$ for $a,b,c = x,y,z$ and $\left\{\bar{P}_{ia}, \bar{P}_{ib}\right\} = -\epsilon_{abc}\bar{P}_{ic}$ for $a,b,c = x,y,z$.\footnote{$\epsilon_{abc}$ is the Levi-Civita symbol}
Since $\P_i$ and $-\bar\P_i$ correspond to the angular momenta, we can define the total angular momentum to be $\P = \sum_{i=1}^N \mathbf{P}_i - \bar{\mathbf{P}}_i$. Similarly, we can identify $\omega_i\P$ and $\omega_i\bar{\P}$ with the magnetic moments, and we can define the total magnetic moment to be $\M = \sum_{i=1}^{N} \omega_i\P_i+\omega_i\bar{\P}_i$. Let us define 
\begin{equation}
  \mathcal{H} \equiv \mathbf{B} \cdot \M + \frac{\mu}{4} \sum_{i,j=1}^N \left(\mathbf{P}_i - \bar{\mathbf{P}}_i\right) \left(\mathbf{P}_j - \bar{\mathbf{P}}_j \right) \left( 1 - \vec{v}_i \cdot \vec{v}_j \right).
\end{equation}
With this Hamiltonian, we can recover \eref{eq:gendiff} from the Poisson bracket formulation of Hamilton's equations, $\dot{f} = \left\{ f, \mathcal{H} \right\}$.

With the formulation in Hamiltonian mechanics, we have identified $\mathcal{H}$ as a conserved quantity. Knowing $\mathcal{H}$ it is also easy to show that the projection of the total angular momentum on the mass vector, $\mathbf{P} \cdot \mathbf{B}$, is conserved. There is, however, one caveat when considering the system as Hamiltonian. Naively, the system seems to be 6N dimensional since there is a three dimensional polarization vector for each neutrino and antineutrino, but as we saw, the phase space is actually only 4N dimensional, and this becomes important when we later interpret the Lyapunov exponents.

\subsection{Two momentum modes}

The specific model, we will consider, has only two momentum states $\vec{p}_1 = - \vec{p}_2$~\cite{Raffelt:2013isa}, and for each of these momentum states, we define the sums $\mathbf{S}_i = \mathbf{P}_i + \bar{\mathbf{P}}_i$ and the differences $\mathbf{D}_i = \mathbf{P}_i - \bar{\mathbf{P}}_i$. In this model the total angular momentum, the total magnetic moment, and the conserved quantities from the Hamiltonian formulation are
\begin{equation}
\begin{aligned}
\P &= \D_{+} \equiv \D_1 + \D_2, \\
\M &= \omega\S_{+} \equiv \omega(\S_1 + \S_2), \\
\mathcal{H} &= \omega \B\cdot \S_+ \mu \D_1\D_2,\\
\B \cdot \P &= \B\cdot \D_{+} = D_{+z}.
\end{aligned}
\end{equation}
The equations of motion can now be found from either $\mathcal H$ or from \eref{eq:gendiff}, and we get
\begin{equation}
  \label{eq:diff}
\begin{aligned}
  \dot{\mathbf{S}}_1 &= \omega \mathbf{B}\times \mathbf{D}_1 + \mu \mathbf{D}_2 \times \mathbf{S}_1 ,\\
  \dot{\mathbf{S}}_2 &= \omega \mathbf{B}\times \mathbf{D}_2 + \mu \mathbf{D}_1 \times \mathbf{S}_2 ,\\
  \dot{\mathbf{D}}_1 &= \omega \mathbf{B}\times \mathbf{S}_1 + \mu \mathbf{D}_2 \times \mathbf{D}_1 ,\\
  \dot{\mathbf{D}}_2 &= \omega \mathbf{B}\times \mathbf{S}_2 + \mu \mathbf{D}_1 \times \mathbf{D}_2 ,
\end{aligned}
\end{equation}
which are the equations we will solve numerically.

In all of the following, we use $\omega = \pm 1$ for the two different mass hierarchies and $\mu=6$ in some arbitrary units, and  we choose to describe a pure electron neutrino beam by $\mathbf{S}_i = 2(\sin(2\theta), 0, \cos(2\theta))$, where we use the mixing angle $\sin(2\theta) = 0.1$. However, before we come to the numerical results, we will briefly review the simplest solutions.

\begin{figure}[tbp]
\center
\includegraphics[width=0.9\columnwidth]{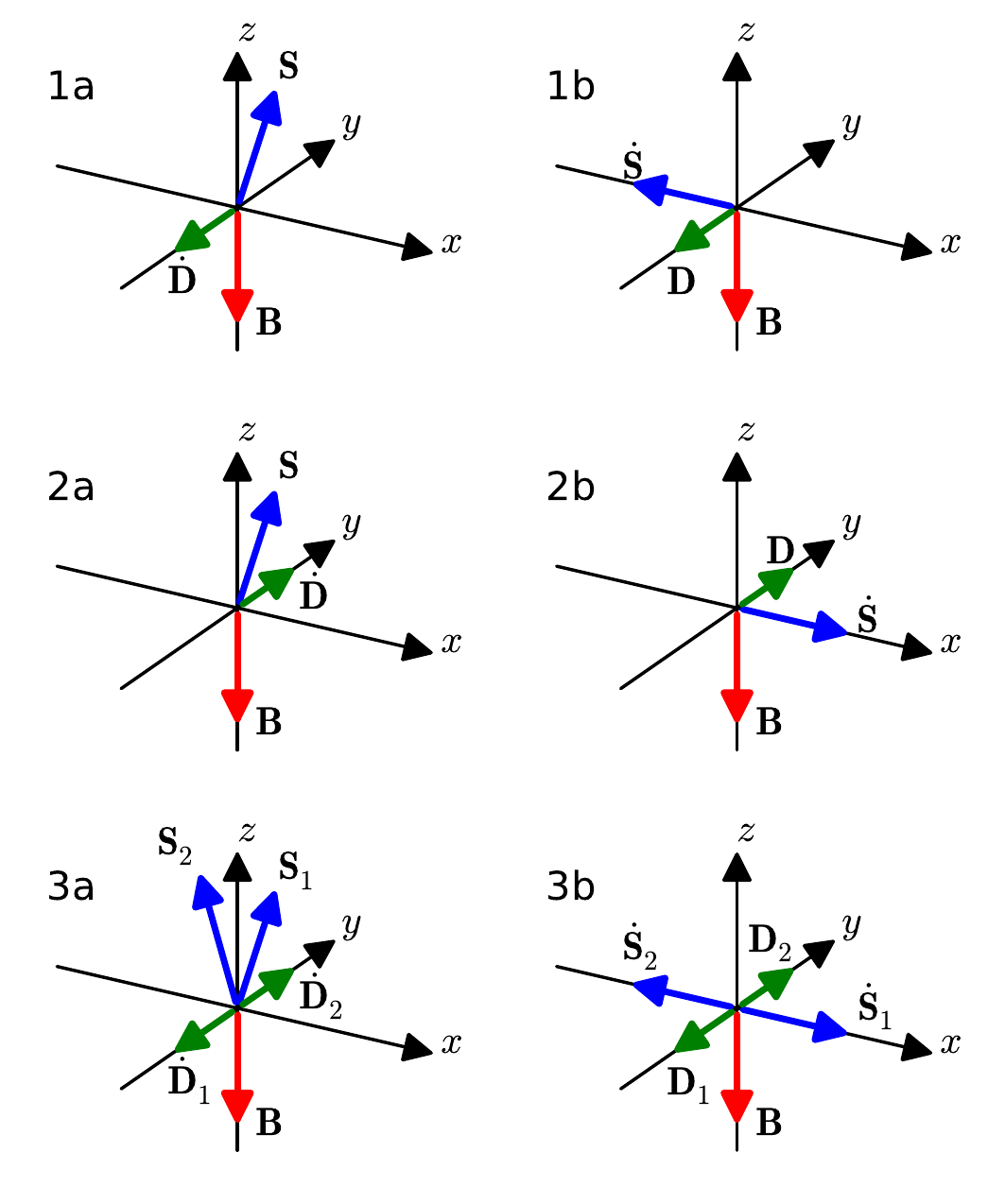}
\caption{Illustrations of different solutions to \eref{eq:diff}. 1a and 1b show the normal hierarchy with symmetrical initial conditions. 2a and 2b show the inverted hierarchy with symmetric initial conditions. 3a and 3b show the normal hierarchy with anti-symmetric initial conditions. 
}
\label{fig:model}
\end{figure}

If we assume that the two momentum states have the same initial conditions, the equations reduce to 
\begin{equation}
  \label{eq:diff1}
\begin{aligned}
  \dot{\mathbf{S}} &= \left(\omega \mathbf{B} - \mu \mathbf{S}\right) \times \mathbf{D} ,\\
  \dot{\mathbf{D}} &= \omega \mathbf{B}\times \mathbf{S}.
\end{aligned}
\end{equation}
This system is equivalent to the isotropic case~\cite{Hannestad:2006nj}, and it supports two different simple solutions depending on the sign of $\omega$, assuming that $\mu>|\omega|$ which corresponds to neutrino-neutrino interactions dominating vacuum oscillations. For the normal hierarchy where $\omega > 0$, a small initial $S_x$-value will make $D_y$ negative which in turn decreases the value of $S_x$. This results in an oscillatory motion around the $z$-axis for $\S$ and oscillations on the $y$-axis for $\D$. Since this solution only deviates slightly from the initial state, we will call it the stationary solution.

For the inverted hierarchy where $\omega < 0$, the same initial condition will make $D_y$ positive. The positive value of $D_y$ will enhance the growth of $S_x$ since $\mu > |\omega|$ making the configuration unstable, and we get a full inversion of the polarization vector. In the literature this type of motion has been compared to an inverted pendulum, and the oscillations are called bipolar since $D_y$ attains a significant value and separates the polarization vectors describing neutrinos and antineutrinos~\cite{Hannestad:2006nj}. 

Let us now go beyond the symmetry assumption and consider the two momentum states separately in the normal hierarchy. For an anti-symmetric initial condition where all components are zero except for $S_{1x} = - S_{2x} > 0$ and $S_{1z}=S_{2z}$, we see that $D_{1y}$ will become negative whereas $D_{2y}$ will become positive. Since $\mu > |\omega|$, we find that $D_{2y}$ dominates $\dot{\S}_1$ while $D_{1y}$ dominates $\dot{\S}_2$ making both unstable as it happened in the inverted hierarchy for the symmetric initial conditions. All of this results in bipolar oscillations as it did for the inverted hierarchy with symmetric initial conditions. In a similar way one can see that the inverted hierarchy will give an approximately stationary solution with the anti-symmetric initial conditions.

While these simple trajectories are solutions to the system, we will also investigate how modifications to the initial conditions turn out to give much more complicated trajectories in polarization space. Before we come to that, we will briefly review Lyapunov exponents and covariant Lyapunov vectors.



\section{Concepts of Lyapunov analysis}
\label{sec:lyalight}

The aim of  Lyapunov analysis is to quantify the divergence of initially nearby solutions of a differential equation.
We will do this by solving the differential equations numerically and obtain a trajectory while we simultaneously consider infinitesimal perturbations and investigate how they grow and shrink.
The primary tool for this is the spectrum of Lyapunov exponents. A Lyapunov exponent, $\lambda_i$, is defined such that the distance between two nearby trajectories on average will grow with the factor $e^{\lambda_i t}$ during the time $t$.
That is, given an infinitesimal perturbation $\mathbf{v}$ we define the Lyapunov exponent as
\begin{equation}
  \label{eq:lambda}
  \lambda = \lim_{t' \rightarrow \infty} \frac{1}{t'} \ln \frac{|| \mathbf{v}(t') ||}{|| \mathbf{v}(t_0) ||} .
\end{equation}
For details on how to evolve $\mathbf{v}$ see Appendix~\ref{sec:lyapunov}.
In a multidimensional phase space, it is of course possible that some nearby trajectories will diverge faster than others, while others again might even converge. We refer to this as unstable and stable directions in phase space, and for $n$ dimensions this gives rise to a spectrum with $n$ Lyapunov exponents; $\lambda_1 \ge \lambda_2 \ge \dots \ge \lambda_n$. A stable direction is associated with a negative Lyapunov exponent while an unstable direction is associated with a positive Lyapunov exponent. If a Lyapunov exponent is zero, we say that the associated direction is marginally stable, and the trajectories are on average not diverging nor converging.
If a system has at least one positive Lyapunov exponent, and the trajectory is bounded, it is said to be chaotic. Furthermore, Lyapunov exponents are a measure of how fast a small perturbation to a given trajectory will grow, and in this sense, a larger Lyapunov exponent will indicate a more unstable system. 

For Hamiltonian systems it is possible to prove that the Lyapunov spectrum will be symmetric such that $\lambda_1=-\lambda_n$, $\lambda_2=-\lambda_{n-1}$, $\dots$, $\lambda_{n/2} = -\lambda_{n/2+1}$~\cite{Benettin1980}, and it can furthermore be shown that each conserved quantity will give rise to two vanishing Lyapunov exponents~\cite{Ginelli2013}. 

Several different ways have been used to associate a direction to the Lyapunov exponent, and we will use the term Lyapunov vector to cover all of them. 
The Lyapunov vector we will mainly use is called the covariant Lyapunov vector, and it is quite conceptually intuitive. The idea of the covariant Lyapunov vector is to generalize the concept of normal modes for a stationary solution to arbitrary trajectories. To do this, we require that the $i$'th covariant Lyapunov vector, $\g_i$, expands with the rate $\lambda_i$ when the differential equations are evolved forward in time. Similarly, it must contract with the rate $-\lambda_i$ if the differential equations are evolved backwards in time. This also means that evolving a covariant Lyapunov vector along the trajectory from $t_1$ to $t_2$ gives the corresponding covariant Lyapunov vector at $t_2$.

A case which needs a special remark, and which will be relevant for the current work, is the degeneracy of Lyapunov exponents. If several Lyapunov exponents have the same value, the associated Lyapunov vectors can be chosen arbitrarily as long as they span the relevant subspace. This means that any linear combination of a set of Lyapunov vectors for degenerate Lyapunov exponents must be considered a Lyapunov vector for those exponents as well.
This is important to notice for the interpretation of the Lyapunov vectors while the generalization from the non-degenerate to the degenerate case is quite trivial for most of the more conceptual considerations.

The neat properties of covariant Lyapunov vectors are not found for other Lyapunov vectors, but these do have other advantages. 
An important disadvantage of the covariant Lyapunov vectors is the inability to decide whether a given direction is divergent or convergent based on a non-zero value of the corresponding vector component. 
Let us look at a three dimensional example to clarify why. Assume, that the spectrum is $\lambda_1 = 1$, $\lambda_2=-1$, and $\lambda_3=-2$. If the corresponding covariant Lyapunov vectors are $\g_1 = \tfrac{1}{\sqrt{2}}(1,1,0)$, $\g_2=\tfrac{1}{\sqrt{5}}(2,0,1)$ and $\g_3=\tfrac{1}{\sqrt{2}}(1,0,-1)$, any vector in the $x,z$-plane can be expressed as a linear combination of $\g_2$ and $\g_3$ proving the vector $(1,0,0)$ to be stable, although $\g_{1,x} \neq 0$ and $\lambda_1 > 0$.
In order to be able to deduce anything about stability from a single component of a Lyapunov vector, it is rather the so-called forward singular vectors, $\mathbf{f}_i$, which should be considered. The forward singular vectors are defined such that all vectors which grow slower than $\lambda_i$ are in the orthogonal compliment of $\mathbf{f}_i$, and they can therefore be obtained by orthogonalization of the covariant Lyapunov vectors starting with the last one. Taking the example from above, the forward singular vectors would be $\mathbf{f}_1 = (0,1,0)$, $\mathbf{f}_2 = \tfrac{1}{\sqrt{2}}(1,0,1)$, and $\mathbf{f}_3 = \tfrac{1}{\sqrt{2}}(1,0,-1)$, and it is clear that $(1,0,0)$ is not an unstable direction.
For most of our results, this is not a concern as the same components are non-zero for covariant Lyapunov vectors and forward singular vectors. When it is a problem, we will discuss the implications.

A more detailed discussion of Lyapunov exponents and Lyapunov vectors is found in Appendix~\ref{sec:lyapunov} while the calculation of both is described in Appendix~\ref{sec:calc}.



\section{Numerical results}
\label{sec:num}

We have solved the equations of motion along with the equations describing the Lyapunov exponents and covariant Lyapunov vectors for four different types of trajectories. However, before we present these results, we will present some of the results which are common for all the different trajectories.

For all the cases, we find numerically that the spectrum of Lyapunov exponents has the form 
\begin{equation}
  (\lambda_1, \lambda_2, 0, 0,0,0,0,0,0,0,-\lambda_2,-\lambda_1)
\end{equation}
for the two beam model.
This is also what one would expect due to the Hamiltonian nature of the system. We expect to see a symmetric spectrum as we already mentioned in \sref{sec:lyalight}, but it is a little more involved to argue for all of the zeros. Since we do our calculations using the polarization vectors, we have $3/2$ times as many variables as the canonical Hamiltonian formulation. Therefore, four of the zeros in the spectrum actually relate to the constraints from the constant lengths of $\P_i$ and $\bar\P_i$ rather than to any conserved quantity. The other four zeros, however, correspond to our two conserved quantities; $\mathcal{H}$ and $\P\cdot\B$. 

The many zeros and the symmetry reduce the Lyapunov spectrum to only two interesting numbers; $\lambda_1$ and $\lambda_2$. 
The values of these depend on $\omega$, $\mu$, and $\theta$, but in this paper our goal is not to map out this dependence. We would also like to remark that we do not need to know the values of $\lambda_i$ with very high precision. It is the order of magnitude we are interested in, and therefore it is not crucial to have a very stringent error estimate either.

Our calculated Lyapunov exponents are seen in \tref{tab:LE}, where we show $\lambda_1$ and $\lambda_2$. We also give an estimate of the uncertainty on our numbers, but note that these are not stringent standard deviations due to some issues with correlated data which we discuss further in Appendix~\ref{sec:calc}. 

For all the cases we have studied, the leading Lyapunov exponent is positive, indicating chaotic behavior. This might seem strange for the stationary and periodic orbits, but for these orbits it is merely a statement of instablity.

\begin{table*}[tbp]
\begin{tabular}{l@{\quad} r@{\quad} r@{\quad} r@{\quad} r@{\quad} r@{\quad} r@{\quad} r@{\quad} r@{\quad} r@{\quad} r@{\quad} r@{\quad} r}
& \multicolumn{2}{c}{No $\delta$ added} & \multicolumn{2}{c}{$\delta S_z$} & \multicolumn{2}{c}{$\delta S_y$} \\
 & \multicolumn{1}{c}{$S_{1x} = -S_{2x}$} & \multicolumn{1}{c}{$S_{1x} = S_{2x}$} & \multicolumn{1}{c}{$S_{1x} = -S_{2x}$} & \multicolumn{1}{c}{$S_{1x} = S_{2x}$} & \multicolumn{1}{c}{$S_{1x} = -S_{2x}$} & \multicolumn{1}{c}{$S_{1x} = S_{2x}$} \\\hline
$\lambda_{1,\mathrm{NH}}$ & $0.99697 \pm 2\cdot10^{-6}$ & $3.3124 \pm 3\cdot10^{-5}$ & $1.19 \pm 0.02$ & $1.234 \pm 0.005$ & $0.97 \pm 0.01$ & $0.973 \pm 0.004$ \\
$\lambda_{2,\mathrm{NH}}$ & $0.5448 \pm 6\cdot10^{-6}$ & $3.3054 \pm 3\cdot10^{-5}$ & $0.76 \pm 0.02$ & $0.794 \pm 0.004$ & $0.53 \pm 0.01$ & $0.521 \pm 0.005$ \\
$\lambda_{1,\mathrm{IH}}$ & $3.3124 \pm 6\cdot10^{-6}$ & $0.5448 \pm 2\cdot10^{-5}$ & $0.76 \pm 0.02$ & $0.753 \pm 0.02$ & $0.68 \pm 0.03$ & $0.720 \pm 0.005$ \\
$\lambda_{2,\mathrm{IH}}$ & $3.3026 \pm 6\cdot10^{-6}$ & $0.0006 \pm 0.0003$ & $0.045 \pm 0.003$ & $0.054 \pm 0.002$ & $0.062 \pm 0.003$ & $0.082 \pm 0.002$ \\
\end{tabular}
\caption{Lyapunov exponents for the normal hierarchy (NH) and the inverted hierarchy (IH) were
 calculated as described in Appendix~\ref{sec:calc}.
}
\label{tab:LE}
\end{table*}

\begin{figure}[tbp]
\center
\includegraphics[width=\columnwidth]{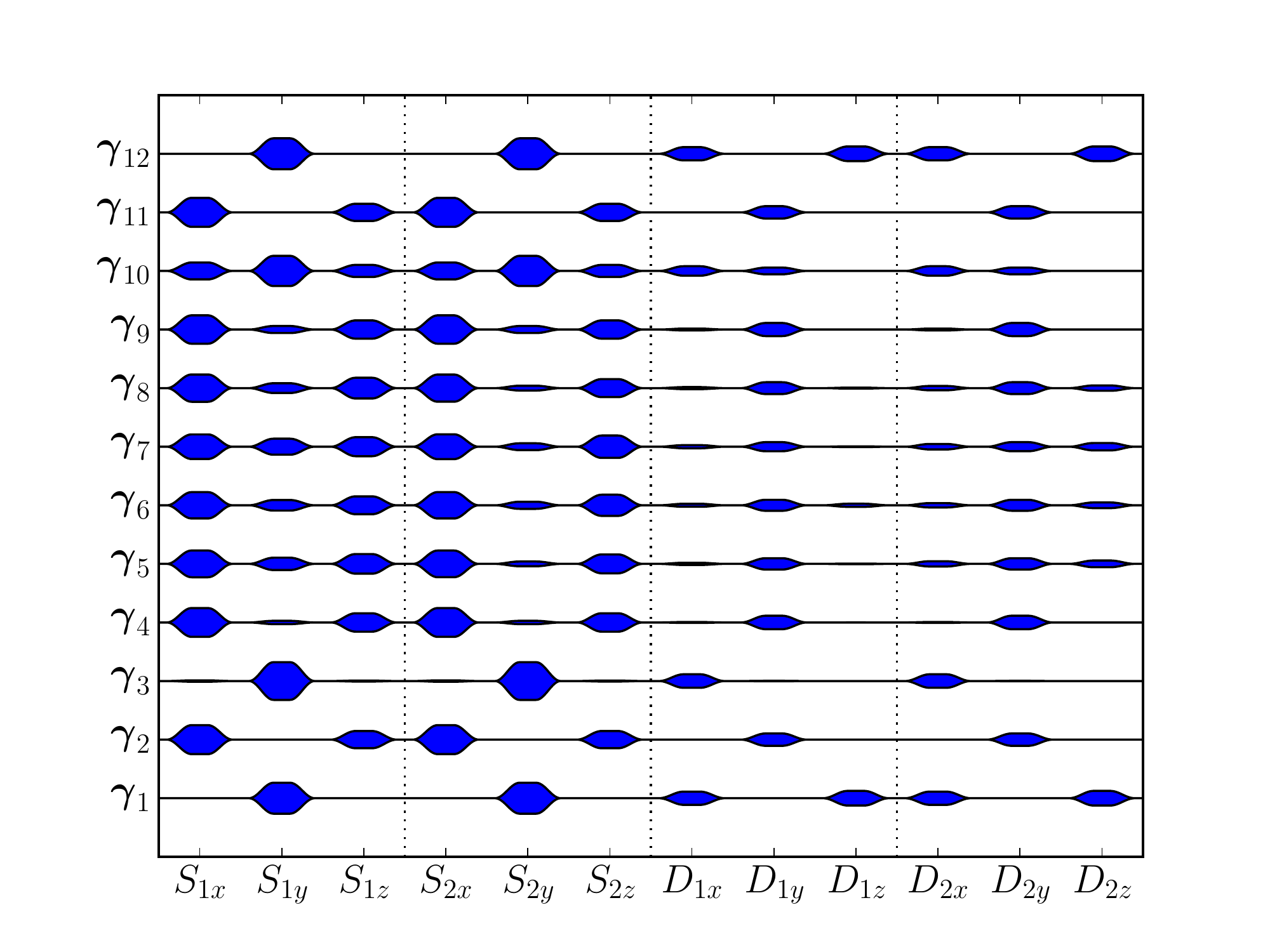}
\caption{All covariant Lyapunov vectors for the normal hierarchy with no modification added to the initial conditions and $S_{1x} = -S_{2x}$ which is the bipolar case. The average magnitude of each component is shown for every vector. The height of the colored area shows how much of the component given on the first axis is present in the vector given on the second axis. e.g. $\g_1$ has equally large components in the $S_{1,y}$- and $S_{2,y}$-directions while the components of $\g_1$ in the $S_{1,x}$-, $S_{1,z}$-, $S_{2,x}$-, and $S_{2,z}$-directions are zero.}
\label{fig:vec_nh_no_p}
\end{figure}

An example of the covariant Lyapunov vectors is shown in \fref{fig:vec_nh_no_p}.  For each coordinate, the average magnitude of that component is shown for every vector. We have computed the covariant Lyapunov vectors for 100000 time steps, but in order to ensure that the computation have actually converged both forward and backward, we skip the first and last $20\%$ when doing the averages.\footnote{We have also tried to skip $40\%$ which gives the same result, so $20\%$ is sufficient to ensure convergence.} From the Figure it is clear that the first covariant Lyapunov vector ($\g_1$) and the last ($\g_{12}$) as well as the second ($\g_2$) and the second last ($\g_{11}$) point in similar directions. We find this to be the case for all our calculations, so we will only be interested in $\g_1$ and $\g_2$ from now on.



\subsection{Stationary solutions}
\label{subsec:stationary}

\begin{figure}[tbp]
\center
\includegraphics[width=\columnwidth]{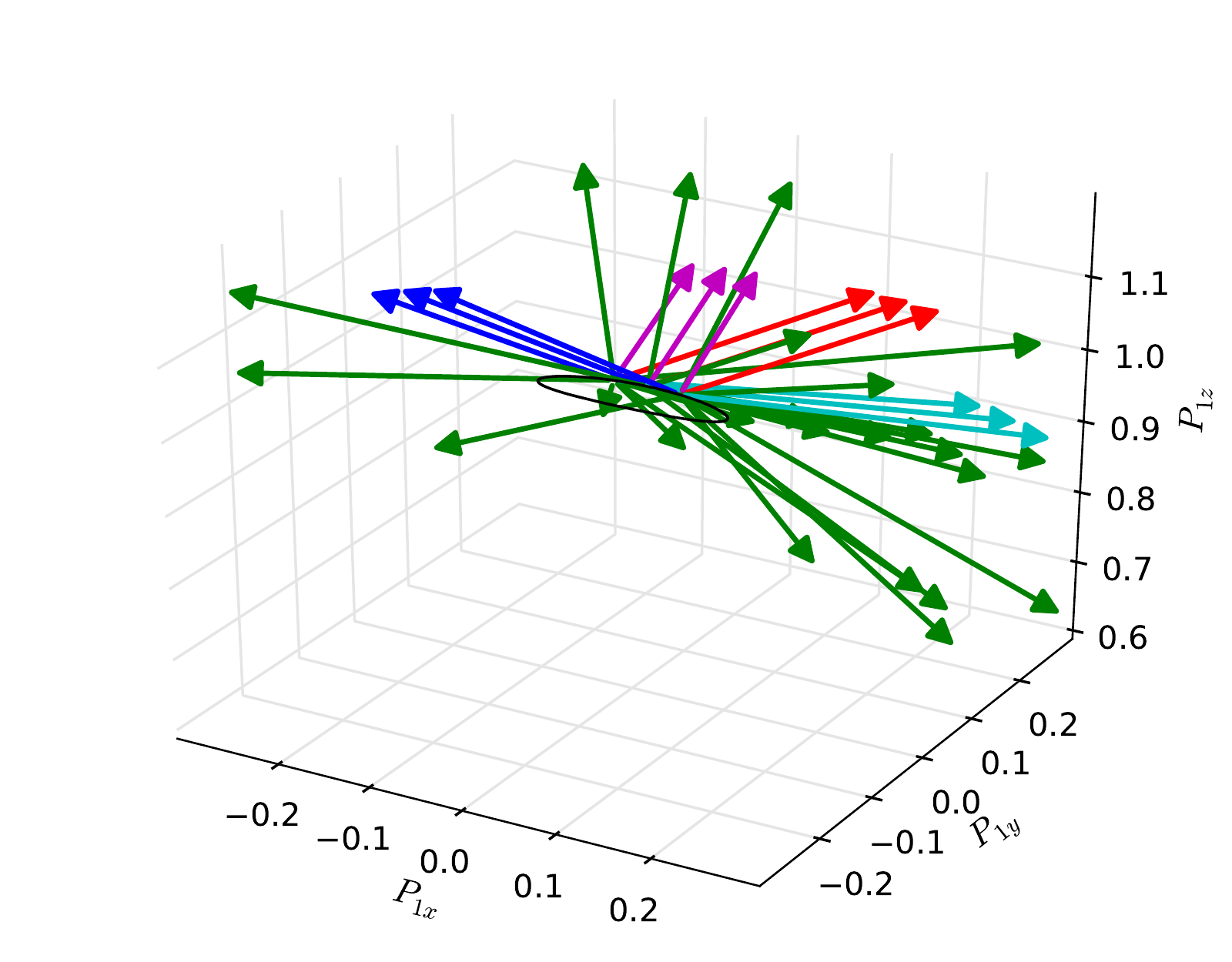}
\caption{The trajectory for the normal hierarchy in the stationary case projected on $\P_1$ is shown in black. Three sets of covariant Lyapunov vectors are also shown. Blue is $\g_1$, red is $\g_2$, greens are $\g_3$ to $\g_{10}$, magenta is $\g_{11}$, and cyan is $\g_{12}$.
}
\label{fig:traj_stat_clv}
\end{figure}

The simplest trajectories, we will consider, are the stationary solutions where the polarization vectors stay very close to the z-axis as seen in \fref{fig:traj_stat_clv}. Although we call this the stationary case, we must remember that the solution is only approximately stationary, and it turns out that the small deviation from a genuinely stationary solution will effect some of the quantities we calculate.

\begin{figure}[tbp]
\center
\includegraphics[width=\columnwidth]{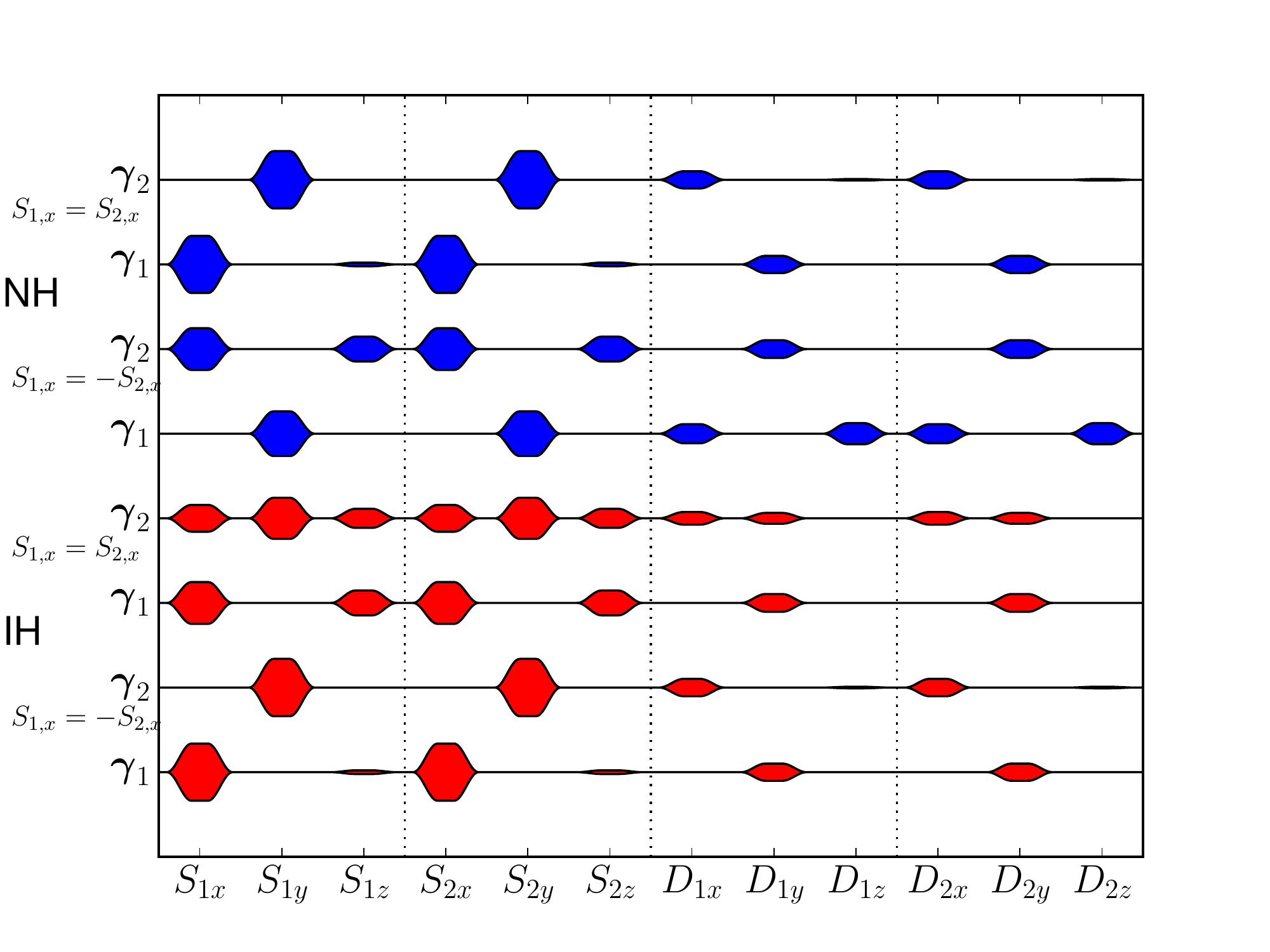}
\caption{Covariant Lyapunov vectors for the static case (two upper and two lower vectors) and the bipolar case (central four vectors) in the $\mathbf{S}_{1,2}, \mathbf{D}_{1,2}$ coordinates. The average magnitude of each component is shown for every vector. Consult \fref{fig:vec_nh_no_p} for notes on how to read the figure.}
\label{fig:vec_no12}
\end{figure}

\begin{figure}[tbp]
\center
\includegraphics[width=\columnwidth]{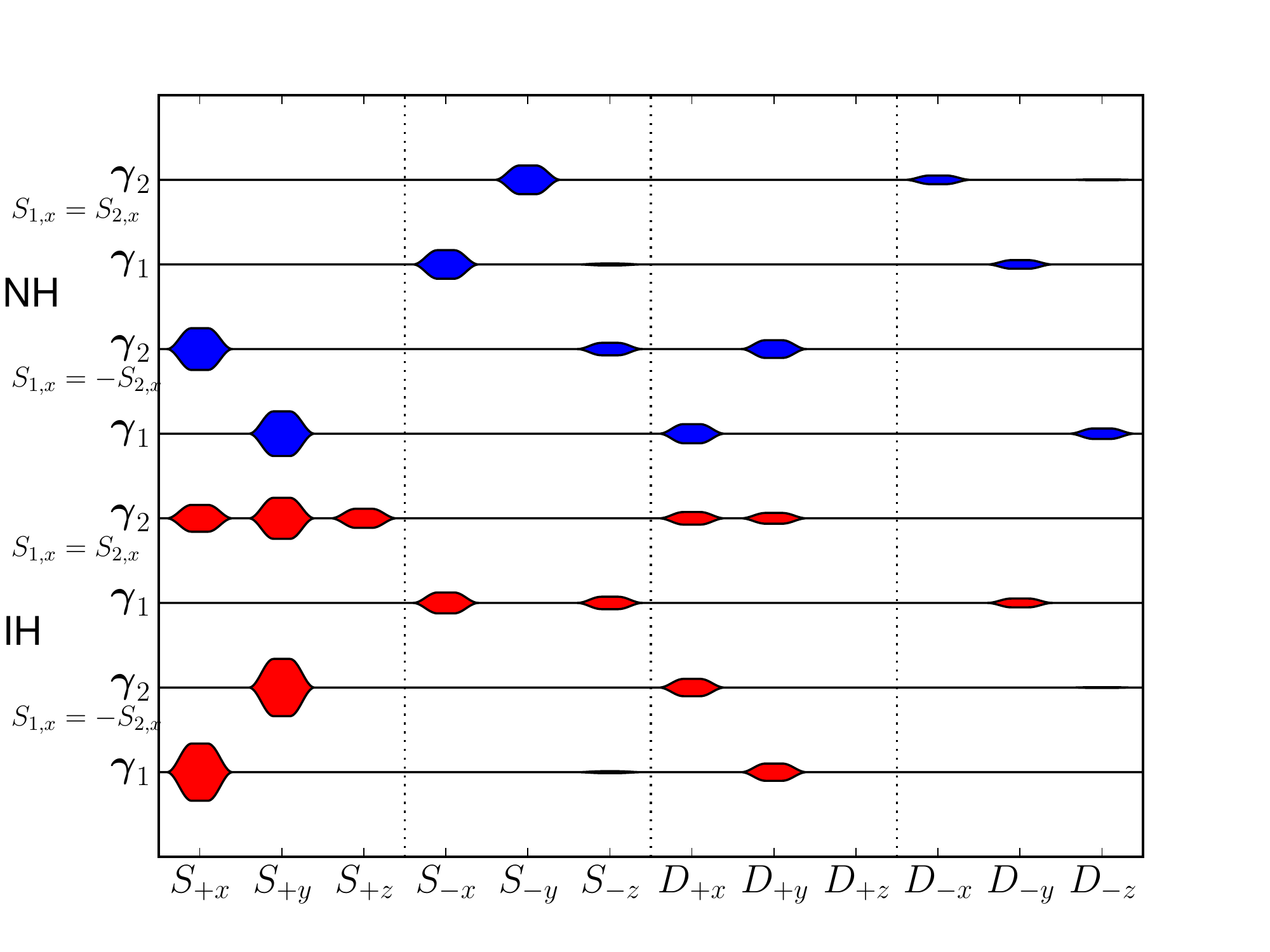}
\caption{Covariant Lyapunov vectors for the static case (two upper and two lower vectors) and the bipolar case (central four vectors) in the $\mathbf{S}_{\pm}, \mathbf{D}_{\pm}$ coordinates. The average magnitude of each component is shown for every vector. Consult \fref{fig:vec_nh_no_p} for notes on how to read the figure.}
\label{fig:vec_nopm}
\end{figure}

For the stationary cases (that is $S_{1x} = S_{2x}$ for the normal hierarchy and $S_{1x} = -S_{2x}$ for the inverted hierarchy), we find that $\lambda_1 \approx \lambda_2 = 3.31$. 
This means that any linear combination of $\g_1$ and $\g_2$ should be considered a covariant Lyapunov vector for the highest Lyapunov exponent. 
A small sample of covariant Lyapunov vectors is shown in \fref{fig:traj_stat_clv}, but it is hard to find any structures when depicting them in this way. Instead we will consider averages of the lengths of each component over time.
These averages can be seen in \fref{fig:vec_no12} for all the perfectly symmetric and anti-symmetric cases, and due to the symmetry between the two momentum states, we also transform the vectors to the $\{+,-\}$-basis where $\mathbf{S}_\pm = \mathbf{S}_1 \pm \mathbf{S}_2$ and $\mathbf{D}_\pm = \mathbf{D}_1 \pm \mathbf{D}_2$ in \fref{fig:vec_nopm}. 
The stationary case is the lower two and the upper two vectors in each figure. In the $\{1,2\}$-basis, the two sets of vectors look very similar, but in the $\{+,-\}$-basis, the normal hierarchy has only a $S_-$ and $D_-$ component while the inverted hierarchy has only a $S_+$ and $D_+$ component. It is also possible to get some analytical insight since the covariant Lyapunov vectors coincide with the normal modes of an ordinary stability analysis when the trajectory is stationary.

Inspired by the stability analysis Duan~\cite{Duan:2013kba} did on the two beam model, we transform \eref{eq:diff} to the $\{+,-\}$-basis:
\begin{equation}
\begin{aligned}
  \dot{\mathbf{S}}_+ &= \omega \mathbf{B}\times \mathbf{D}_+ + \frac{\mu}{2} \mathbf{D}_+ \times \mathbf{S}_+ - \frac{\mu}{2} \mathbf{D}_- \times \mathbf{S}_- \\
 &\approx (\omega+\mu) \mathbf{B} \times \mathbf{D}_+ ,\\
  \dot{\mathbf{S}}_- &= \omega \mathbf{B}\times \mathbf{D}_- + \frac{\mu}{2} \mathbf{D}_+ \times \mathbf{S}_- - \frac{\mu}{2} \mathbf{D}_- \times \mathbf{S}_+ \\
 &\approx (\omega-\mu) \mathbf{B} \times \mathbf{D}_-,\\
  \dot{\mathbf{D}}_+ &= \omega \mathbf{B}\times \mathbf{S}_+,\\
  \dot{\mathbf{D}}_- &= \omega \mathbf{B}\times \mathbf{S}_- + \mu \mathbf{D}_+ \times \mathbf{D}_-\\
 &\approx \omega \mathbf{B}\times\mathbf{S}_- ,
\end{aligned}
\end{equation}
where we have used the approximations $\S_+/2 \approx \mathbf{S}_1 \approx \mathbf{S}_2 \approx -\mathbf{B}$, and $\mathbf{D}_{+}$, $\D_-$, and $\S_-$ are small, so some quadratic terms can be neglected.

From these equations it is clear that $+$ and $-$ decouple, and we find 
\begin{equation}
  \begin{gathered}
  \ddot{\mathbf{S}}_+ \approx (\omega + \mu) \omega \mathbf{B} \times (\mathbf{B} \times \mathbf{S}_+) = -\omega(\omega+\mu)\mathbf{S}_+,\\
  \ddot{\mathbf{D}}_+ \approx -\omega(\omega + \mu) \mathbf{D}_+,\\
  \ddot{\mathbf{S}}_- \approx -\omega(\omega-\mu)\mathbf{S}_-,\\
  \ddot{\mathbf{D}}_- \approx -\omega(\omega - \mu) \mathbf{D}_-.\\
\end{gathered}
\end{equation}
For the normal hierarchy, we get the solutions
\begin{equation}
  \begin{gathered}
  \mathbf{S}_+ = \mathbf{a}_1 e^{\pm itk_+},\quad \mathbf{D}_+ = \mathbf{a}_2 e^{\pm i tk_+}, \quad k_+ = \sqrt{\omega(\mu+\omega)}\\ \mathbf{S}_- = \mathbf{a}_3 e^{\pm tk_-},\quad \mathbf{D}_- = \mathbf{a}_4 e^{\pm tk_-},\quad k_- = \sqrt{\omega(\mu-\omega)}.
\end{gathered}
\end{equation}
This suggest that $\lambda_1 = \lambda_2 = -\lambda_{11} = -\lambda_{12}$, and that the covariant Lyapunov vectors should point towards $\mathbf{S}_-$ and $\mathbf{D}_-$ in the normal hierarchy as it is seen for the two upper vectors in \fref{fig:vec_nopm}. $S_{-z}$ and $D_{-z}$ are almost zero since both their derivatives are approximated by $\propto \mathbf{B}\times\mathbf{X}$. As $\mathbf{B} = (0,0,-1)$, the derivative in the $z$-direction is zero. The small deviations from zero are due to the fact that the simulated system is not perfectly stationary. If we set $\sin^2(2\theta) = 0$, we find the two $z$-components to be exactly zero.

A similar analysis can be done for the inverted hierarchy. Here the sign of $\omega$ is opposite, and the solutions are
\begin{equation}
  \begin{gathered}
  \mathbf{S}_+ = a_1 e^{\pm tk_+},\quad \mathbf{D}_+ = a_2 e^{\pm tk_+}, \quad k_+ = \sqrt{-\omega(\mu+\omega)}\\ \mathbf{S}_- = a_3 e^{\pm itk_-},\quad \mathbf{D}_- = a_4 e^{\pm itk_-},\quad k_- = \sqrt{-\omega(\mu-\omega)}.
  \end{gathered}
\end{equation}
Again this is consistent with the numerical result in \fref{fig:vec_nopm}.

From a more intuitive point of view, we notice that we have $S_{1x} = S_{2x}$ in the initial condition for the normal hierarchy, and this turns out to hold true for all times. 
Therefore, any perturbation acting symmetrically on $S_{1x}$ and $S_{2x}$ or $S_{1y}$ and $S_{2y}$  will conserve the symmetry of the system. On the contrary perturbations acting anti-symmetrically will break the symmetry. These two cases correspond to perturbations in $S_{+x}$ and $S_{+y}$ versus $S_{-x}$ and $S_{-y}$ respectively, so $\g_1$ and $\g_2$ must point in the directions of $S_{-x}$ and $S_{-y}$ as we also find. In the inverted hierarchy $S_{1x} = -S_{2x}$, and anti-symmetric perturbations will conserve the symmetry whereas symmetric perturbations will break the symmetry of the system, so $\g_1$ and $\g_2$ must point in the directions of $S_{+x}$ and $S_{+y}$. All of this is consistent with the numerical and analytical results.

With a better understanding of the covariant Lyapunov vectors, we will now consider the stability of the system. For the positive and negative Lyapunov exponents, we note that the stable and unstable directions in polarization space are coincident according to the covariant Lyapunov vectors. This is also the result in our algebraic analysis where we find both positive and negative exponentials to solve the differential equations. Consequently, it is impossible to find a set of converging solutions as the diverging solution will always dominate. 
Apart from the positive and negative Lyapunov exponents, we found eight vanishing exponents which we can interpret in terms of constraints and conserved quantities. With regard to stability, however, their associated covariant Lyapunov vectors indicate marginally stable directions in which perturbation will neither shrink nor grow on average.
From the intuitive point of view, we find that the marginally stable directions correspond to perturbations which do respect the symmetry of the system. Also, it turns out that all the $z$-components relate to vanishing Lyapunov exponents in the perfectly stationary case. For the more realistic case where $\theta \neq 0$, however, we find that the directions of $S_{-z}$ and $D_{-z}$ are unstable as well.

A more physical interpretation of the perturbations is slightly hampered by the fact that we have chosen the $z$-axis along the direction of $\mathbf{B}$ and not in the direction of the pure flavor state. In order to recover the more interpretable coordinate system, we would have to rotate all the vectors with the angle $\theta$ in the $(x,z)$-plane. This means that whenever we encounter an $x$-component it actually contains a little of the flavor $z$-component and vice versa.
Apart from this minor complication, a perturbation in any $z$-coordinate corresponds to a perturbation in the flavor content of the neutrinos while a perturbations in the $x$- and $y$-coordinates correspond to perturbation in the phase of the neutrino oscillations.
From this point of view, a perfectly stationary system with vanishing mixing angle would be marginally stable towards perturbations in the flavor content but unstable towards anything that could shift the phase.
For the case of a non-vanishing mixing angle, the inverted hierarchy will be unstable towards flavor perturbations although the components of the covariant Lyapunov vectors are small. The normal hierarchy will also be unstable towards anti-symmetric perturbations but marginally stable towards symmetric flavor perturbations as it is seen in \fref{fig:vec_nopm}.



\subsection{Bipolar solutions}
\label{subsec:bipolar}

\begin{figure}[tbp]
\center
\includegraphics[width=\columnwidth]{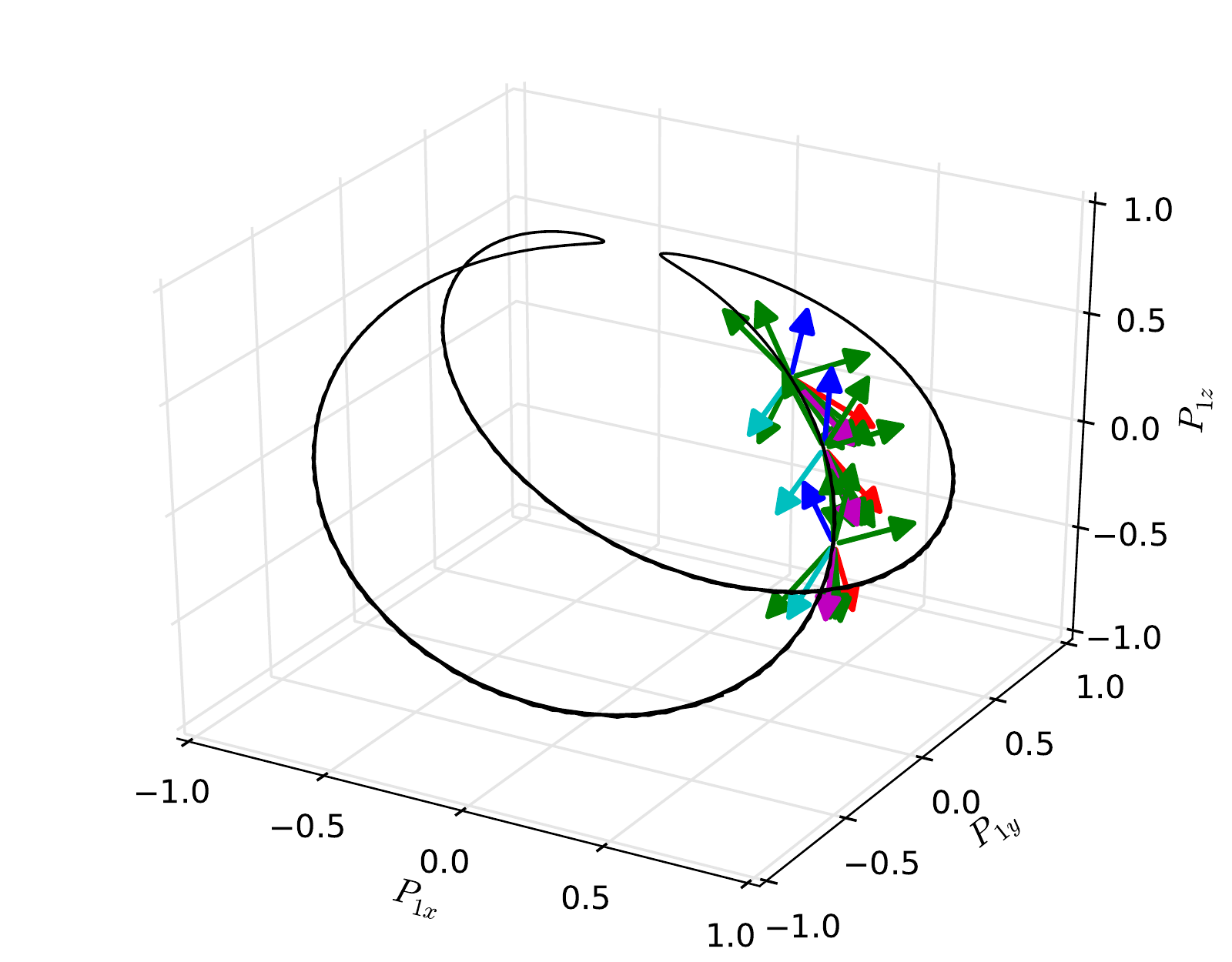}
\caption{The trajectory for the normal hierarchy in the bipolar case projected on $\P_1$. Three sets of covariant Lyapunov vectors are also shown. Blue is $\g_1$, red is $\g_2$, greens are $\g_3$ to $\g_{10}$, magenta is $\g_{11}$, and cyan is $\g_{12}$.}
\label{fig:traj_bip_clv}
\end{figure}

For the bipolar case, we get the well known periodic solutions where the polarization vectors oscillate from $P_z>0$ to $P_z<0$~\cite{Hannestad:2006nj, Raffelt:2013isa, Duan:2005cp} as seen in \fref{fig:traj_bip_clv}. In the normal hierarchy, this solution is obtained when $S_{1x} = -S_{2x}$, and in the inverted hierarchy, we find it for $S_{1x} = S_{2x}$ as we described in \sref{sec:model}.
For the Lyapunov exponents, we get the values $\lambda_1 = 0.997$ and $\lambda_2 = 0.545$ in the normal hierarchy, while the inverted hierarchy gives the values $\lambda_1 = 0.545$ and $\lambda_2 = 0.0006 \approx 0$.
This is remarkable since it suggests that $\lambda_{2,\NH} = \lambda_{1,\IH}$, and it shows the existence of two more vanishing Lyapunov exponents for the inverted hierarchy. The vanishing Lyapunov exponents could suggest that there is another conserved quantity, but since they only vanish for the symmetric and not for the anti-symmetric initial conditions, it is probably rather an artifact of the specific bipolar solution.
When the values are compared to the stationary case, we note that the first two Lyapunov exponents are not degenerate any more and that $\lambda_{\textrm{bipolar}} < \lambda_{\textrm{stationary}}$.

Regarding stability analysis, this suggests the stationary cases to be more unstable than the bipolar ones. 
This also means that a perturbation in the normal hierarchy will need three times longer to grow by the same factor in the bipolar case than in the stationary cases. For the inverted hierarchy, it will need six times as long. 
If this result transfers to real physical systems, it can have an important impact since these perturbations will grow only while $\mu$ is large. In a supernova, $\mu$ becomes smaller as you go away from the center of the supernova, and in the early universe, $\mu$ decays with the expansion of the universe. This limits the time a perturbation has to grow, and the value of the Lyapunov exponents can thus determine if a small perturbation becomes large and makes the trajectory non-periodic.

We will now turn to the covariant Lyapunov vectors. Again we see a sample the trajectory in \fref{fig:traj_bip_clv}, but we still find the averages to be more interesting.
When we consider the four central covariant Lyapunov vectors in \fref{fig:vec_no12}, the pattern from the Lyapunov exponents is repeated as $\g_{1,\IH}$ is very similar to $\g_{2,\NH}$. On the other hand, there is no information in $\g_{2,\IH}$ since its Lyapunov exponent is 10 times degenerate.
Going to \fref{fig:vec_nopm}, it is only the perturbations breaking the symmetry which actually grow as we also saw for the stationary solution. We see that the normal hierarchy with the initial condition $S_{1x} = -S_{2x}$ is stable towards anti-symmetric perturbations ($S_{-x}$ and $S_{-y}$) but unstable with regards to symmetric perturbations ($S_{+x}$ and $S_{+y}$). In the same way, the inverted hierarchy with $S_{1x} = S_{2x}$ is stable towards perturbations in $S_{+x}$ and $S_{+y}$ but unstable with regards to perturbations in $S_{-x}$ and $S_{-y}$.

As for the stationary case, we can interpret the missing components of $\g_1$ and $\g_2$ (in the normal hierarchy) as directions in polarization space far more stable against perturbations than the other directions. We see that perturbations in $S_{1y}$, $S_{2y}$, $D_{1x}$, $D_{1z}$, $D_{2x}$, and $D_{2z}$ are marginally stable in the inverted hierarchy, while their exponential growth is approximately twice as fast as that of other perturbations in the normal hierarchy. 

In a physical interpretation, this is interesting since it shows that small symmetric perturbations in the flavor content or the phase will not be important in the inverted hierarchy if the initial conditions are approximately symmetrical. For the normal hierarchy, it is tempting to draw the same conclusion regarding the flavor content, but here we must remember that the non-zero $x$-component also contains some of the flavor $z$-component.



\subsection{Non-periodic solutions}
\label{subsec:chaotic}

While the bipolar case and the stationary case have been studied for about a decade~\cite{Hannestad:2006nj, Duan:2005cp, Duan:2006an, EstebanPretel:2007ec, Fogli:2007bk}, the interest in the chaotic, non-periodic solutions is quite recent~\cite{Raffelt:2013isa,Duan:2013kba}.

\begin{figure}[tbp]
\center
\includegraphics[width=\columnwidth]{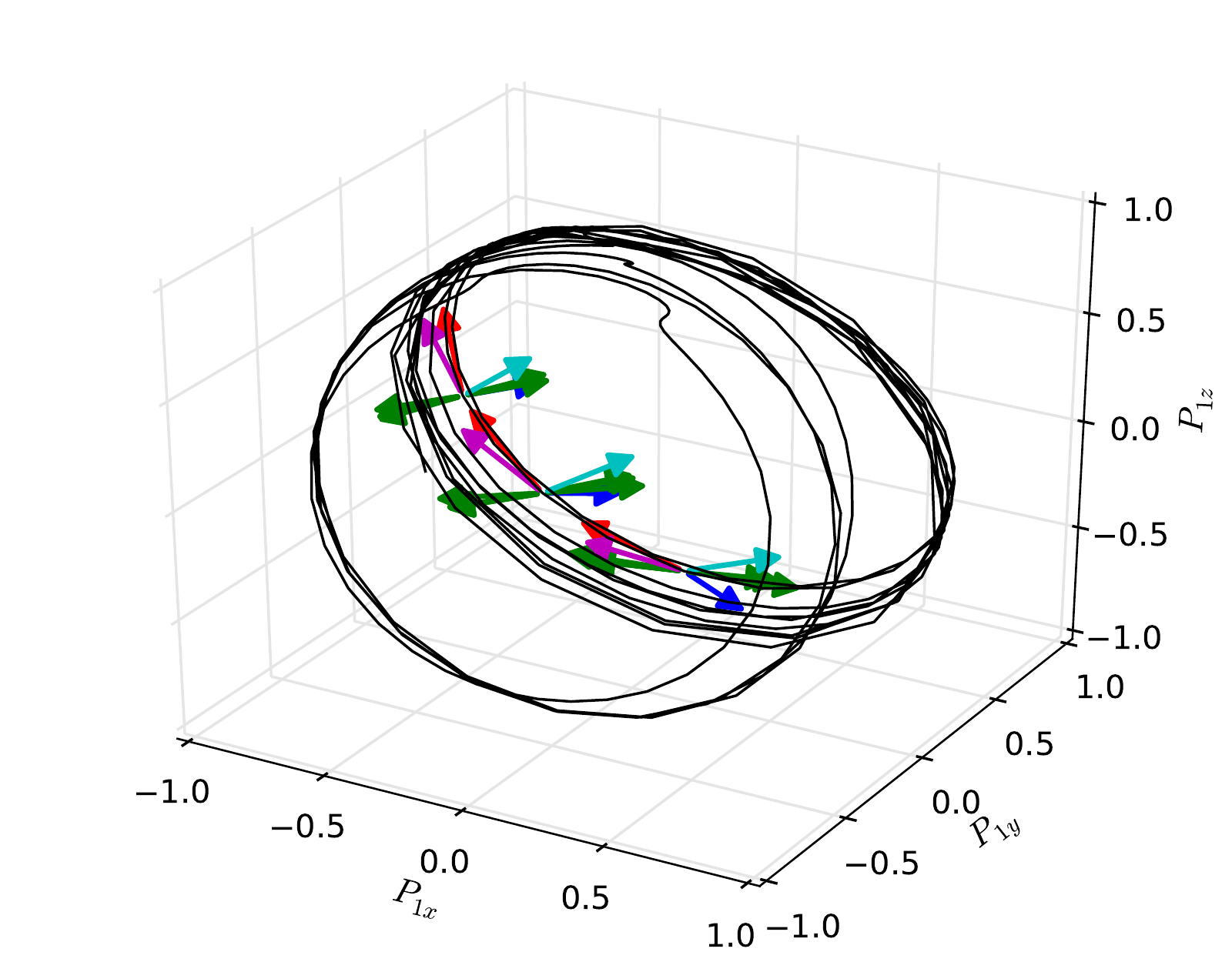}
\caption{The trajectory for the normal hierarchy in the mildly chaotic case projected on $\P_1$. Notice how the trajectory to some degree follows the bipolar solution. The change from $P_{1y}>0$ to $P_{1y}<0$ happens very rarely, but it is seen in this example. Three sets of covariant Lyapunov vectors are also shown. Blue is $\g_1$, red is $\g_2$, greens are $\g_3$ to $\g_{10}$, magenta is $\g_{11}$, and cyan is $\g_{12}$.}
\label{fig:traj_sz_clv}
\end{figure}

\begin{figure}[tbp]
\center
\includegraphics[width=\columnwidth]{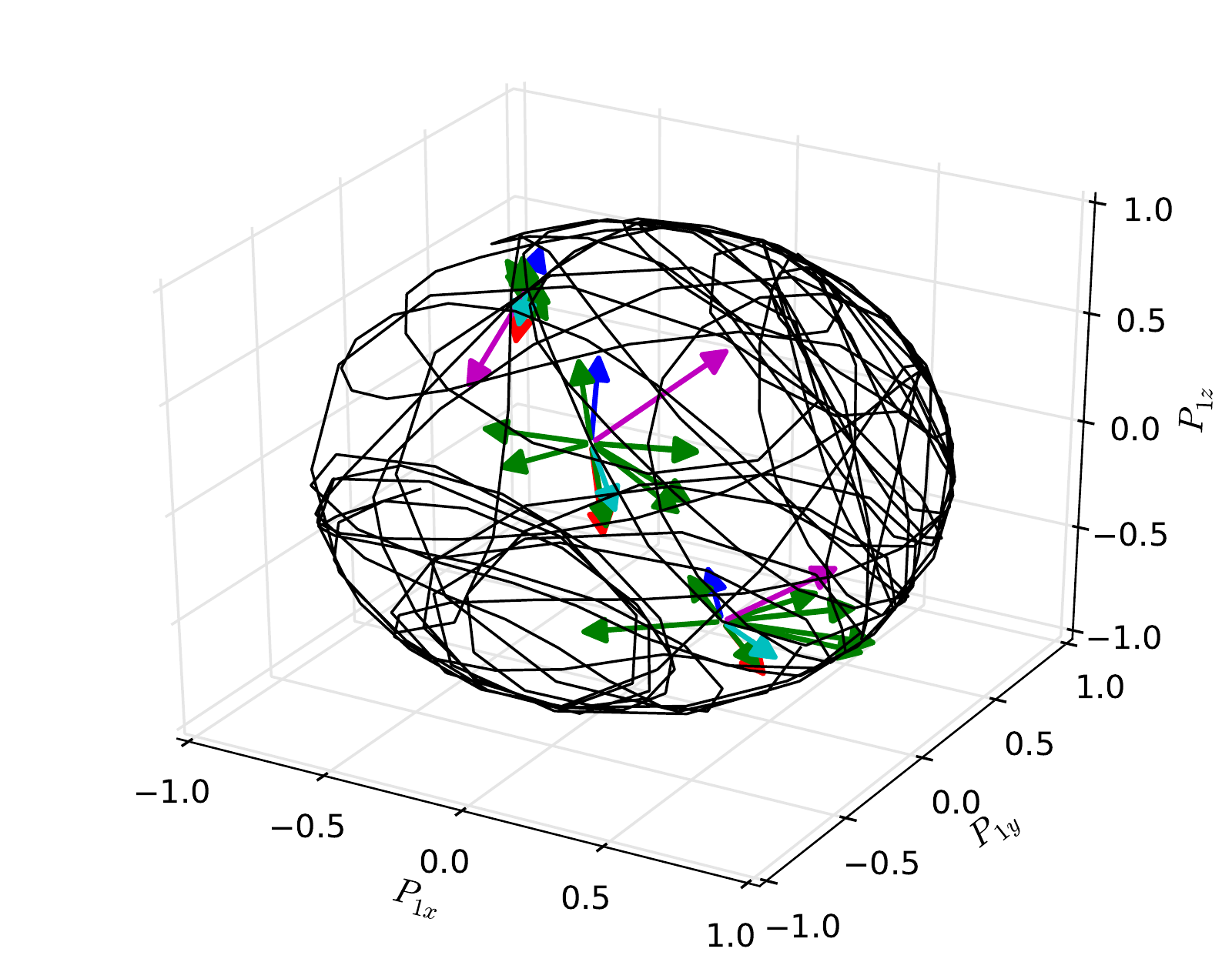}
\caption{The trajectory for the normal hierarchy in the highly chaotic case projected on $\P_1$. Three sets of covariant Lyapunov vectors are also shown. Blue is $\g_1$, red is $\g_2$, greens are $\g_3$ to $\g_{10}$, magenta is $\g_{11}$, and cyan is $\g_{12}$.}
\label{fig:traj_sy_clv}
\end{figure}

The non-periodic solutions we have considered are obtained by taking the initial conditions corresponding to the stationary and bipolar solutions and add $\delta = 2\cdot10^{-3}$ to one of the coordinates. This modification is large enough to make the marginally stable directions non-periodic, but if there is a difference between $S_{1x} = S_{2x}$ and $S_{1x} = -S_{2x}$, we still expect to see it. With this approach, we have found two different types of non-periodicity. For some small modifications of the stationary and periodic cases, we get a trajectory which is not recurrent but stays close to the periodic solution known from the bipolar case as it is seen in \fref{fig:traj_sz_clv}. A property of this group of solutions is that the trajectory stays in the $S_{ix}$-$S_{iz}$ planes and along $D_{iy}$. These coordinates are also the ones where we can add our $\delta$ without making the trajectory even more non-periodic.
If we modify $S_{iy}$, $D_{ix}$, or $D_{iz}$, we get a trajectory which eventually covers the full polarization space fulfilling that $|\mathbf{P}_i|$ and $|\bar{\mathbf{P}}_i|$ are conserved as seen in \fref{fig:traj_sy_clv}. How fast it will deviate significantly from the bipolar oscillations depends on which mass hierarchy we consider, and we will return to this point when discussing the covariant Lyapunov vectors.

When doing the Lyapunov analysis, we have chosen to modify $\delta S_z$ to represent the mildly non-periodic case and modify $\delta S_y$ to represent the most chaotic case. 
Modifying all the other coordinates give results similar to either one or the other. We find the covariant Lyapunov vectors to be very similar within each group while the Lyapunov exponents are within $\sim 20\%$ for each group. 

The first Lyapunov exponents for the non-periodic cases, $\lambda_1$, range from 0.68 to 1.23, so there is no large difference in how fast perturbations grow in the dominantly unstable directions. The second Lyapunov exponents, $\lambda_2$, however, shows a significant difference. In the normal hierarchy, the second Lyapunov exponents range from 0.52 to 0.79, but in the inverted hierarchy, the largest $\lambda_2$ is 0.082. This is not much larger than zero, and it indicates that there are directions which are almost marginally stable in the inverted hierarchy.

Comparing $\lambda_1$ to the stationary and bipolar cases, we find them to be approximately one third of the leading Lyapunov exponents in the stationary cases but quite similar to the bipolar case. As in the bipolar case, this means that a certain perturbation would need about three times longer to affect the solution in the non-periodic cases than it would in the stationary cases.

\begin{figure}[tb]
\center
\includegraphics[width=\columnwidth]{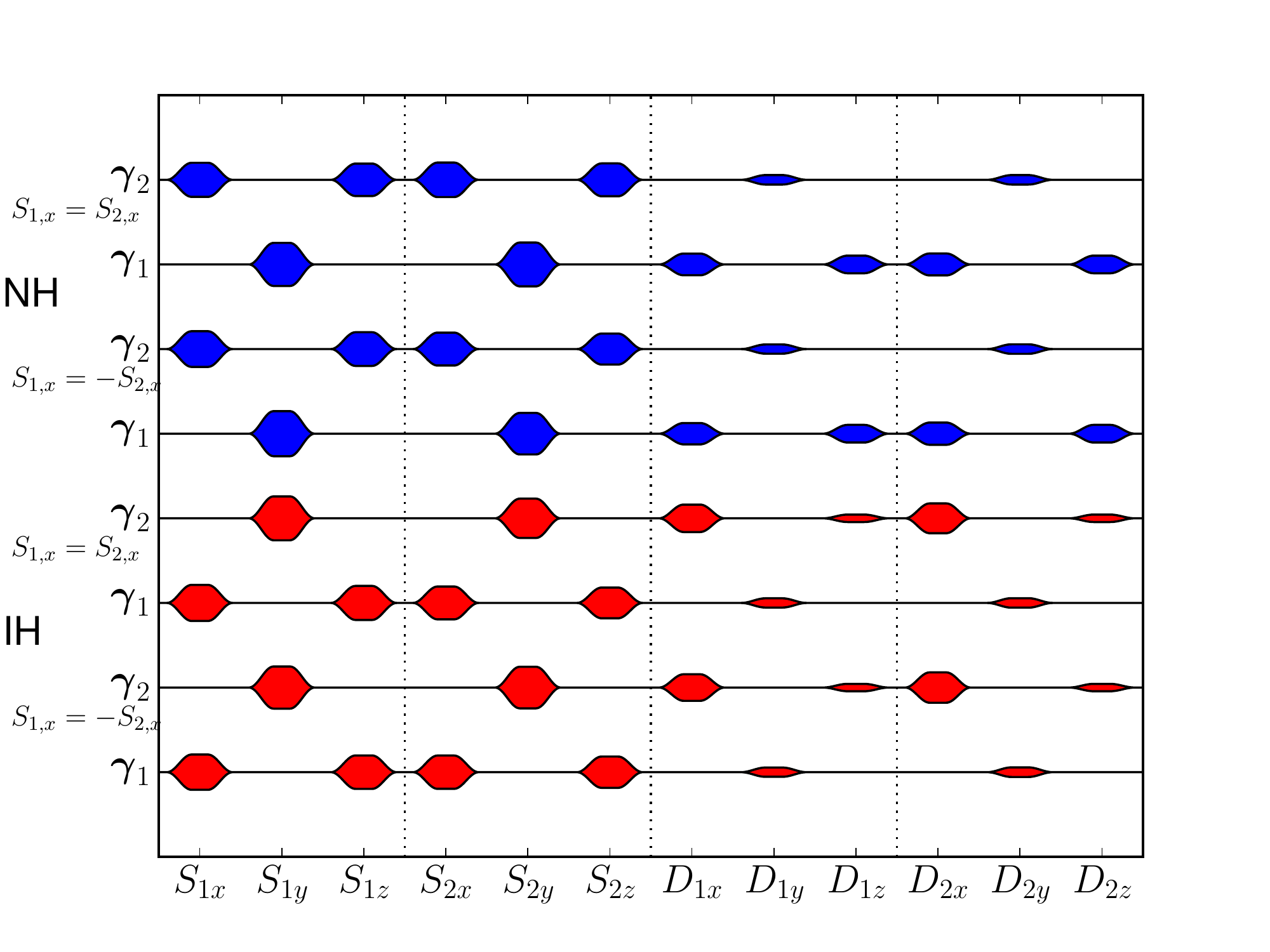}
\caption{Covariant Lyapunov vectors for modifications in the $\delta S_z$ direction in the $\mathbf{S}_{1,2}, \mathbf{D}_{1,2}$ coordinates. The average magnitude of each component is shown for every vector. Consult \fref{fig:vec_nh_no_p} for notes on how to read the figure.}
\label{fig:vec_z12}
\end{figure}

\begin{figure}[tb]
\center
\includegraphics[width=\columnwidth]{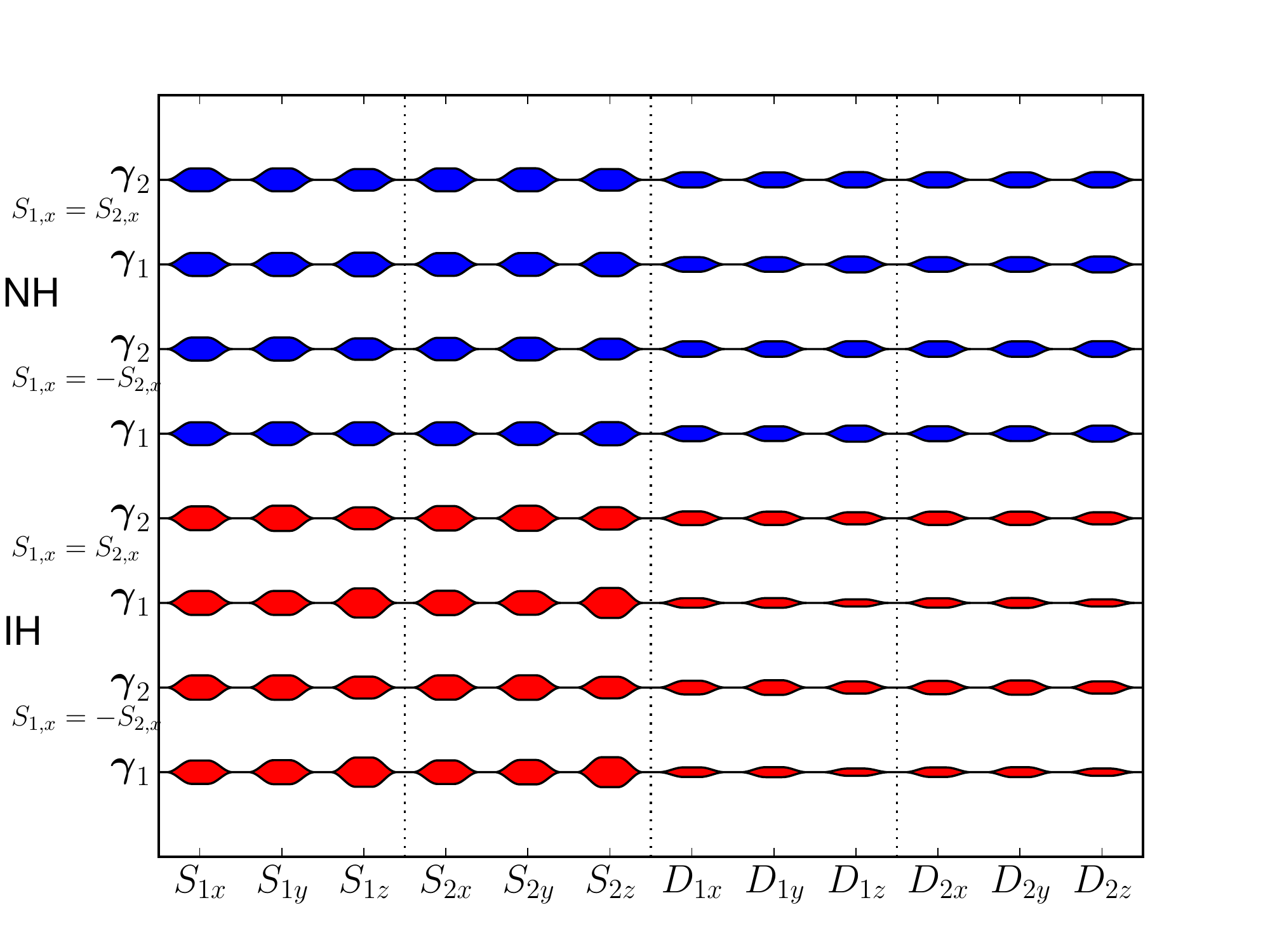}
\caption{Covariant Lyapunov vectors for modifications in the $\delta S_y$ direction in the $\mathbf{S}_{1,2}, \mathbf{D}_{1,2}$ coordinates. The average magnitude of each component is shown for every vector. Consult \fref{fig:vec_nh_no_p} for notes on how to read the figure.}
\label{fig:vec_y12}
\end{figure}

For the non-periodic cases it is even harder to digest the covariant Lyapunov vectors when shown along the trajectory due to its complicated nature. A few examples are shown in \fref{fig:traj_sz_clv} and \fref{fig:traj_sy_clv}, but again we need to consider the averages to find patterns.
The averages of the first and second covariant Lyapunov vectors are seen in \fref{fig:vec_z12} for $\delta S_z$ and in \fref{fig:vec_y12} for $\delta S_y$. As in the previous cases, we have also tried to plot the vectors in the $\{+,-\}$-basis, but it does not provide any additional information except for the fact that the $D_{+z}$-component is zero. This is what we expect for generalized normal modes since $-D_{+z}$ is the projection of the total angular momentum on $\B$ which is a conserved quantity. If we however plot the forward singular vectors in the $\{+,-\}$-basis, we find the $D_{+z}$-component to be non-zero for the $\delta S_y$ cases. For $\delta S_z$ we find the same for $\g_1$ in normal hierarchy and $\g_2$ in inverted hierarchy. This means that a small modification in the $D_{+z}$-direction will lead to a diverging solution. Not that it will diverge in the $D_{+z}$-direction, which is conserved, but the non-linear evolution will transfer the difference to other non-conserved coordinates.

If we now go back to $\delta S_z$ in \fref{fig:vec_z12}, we see some structure.
$\lambda_1$ and $\lambda_2$ are comparable in size and of order one for the normal hierarchy, and the directions associated with $\g_1$ will only diverge a little faster than directions associated with $\g_2$. As a result, the difference between $\g_1$ and $\g_2$ will not have large consequences in the normal hierarchy.
For the inverted hierarchy, on the other hand, $\lambda_2 \sim 0$ while $\lambda_1\sim 1$. This means that directions associated with $\g_2$ will diverge much slower than directions associated with $\g_1$. If we go back and compare the non-zero coordinates of $\g_2$ to the coordinates which give rise to the more chaotic solution exemplified by the $\delta S_y$ case, we find them to be identical. Interestingly enough, we also find these highly non-periodic trajectories to be slower filling out the phase space for the inverted hierarchy than for the normal hierarchy. We believe this to be a remnant from the low value of $\lambda_2$ in the less chaotic $\delta S_z$ case.
For the stability of the system, this vanishing components of $\g_1$ in the inverted hierarchy indicate marginally stable directions. From a physical point of view, however, it is not clear how to perturb only $S_{1x}$ and not $S_{1y}$, so all the phases must be regarded unstable. On the other hand, a flavor perturbation with opposite effects on neutrinos and anti neutrinos would be marginally stable since the $D_{ix}$-components are vanishing as well as the $D_{iz}$-components, and this type of perturbation might be possible.

We will now turn to $\delta S_y$ in \fref{fig:vec_y12} where we see that there is very little information to be gained. All components are present in all vectors. This is understandable since the trajectories cover all of the allowed polarization space which means we average over vectors pointing in all directions.



\section{Conclusions} 
\label{sec:conclusions}

Our analysis has shown a way to generalize the linear stability analysis to periodic and even non-periodic solutions of a set of differential equations.
The Lyapunov exponents quantifies how fast a small perturbation to a known solution can be expected to grow, and the covariant Lyapunov vectors generalize the normal modes from the stationary case and contain information about the dynamics of a given trajectory.

In the simple two beam model we have considered here, we have shown that the stationary case with very little flavor conversion is more unstable than the bipolar flavor changing case since the Lyapunov exponents are larger.
Furthermore the covariant Lyapunov vectors show that some directions are marginally stable, so that perturbations confined to these directions will need a very long time to grow significantly if they will grow at all. This is for example the case for any perturbations in $S_{+z}$ where the flavor content of all neutrinos and anti-neutrinos are perturbed by the same amount.

For the non-periodic trajectories, we have investigated two different types of variations to the stationary and bipolar cases which lead to quite different behavior of the polarization vectors. In the less chaotic case, the trajectory resembles the bipolar solution to some degree, and some of the symmetries in the equations are unbroken.
Again we found that the second Lyapunov exponent was quite small in the inverted hierarchy, and for the less chaotic case this means that perturbations in some directions grow very slowly. In the more chaotic case we struggled to find any structure in the covariant Lyapunov vectors at all, although they do have a tendency to be orthogonal to the polarization vectors which would also be expected since the lengths of the polarization vectors are conserved.

Physically the consequence of a large Lyapunov exponent is that a small perturbation introduced by a thermal fluctuation, an anisotropy, or an inhomogeneity might grow very fast. This is the case if the perturbation has a component of the Lyapunov vector corresponding to the large Lyapunov exponent.
Therefore, the difference we have found between the Lyapunov exponents for the inverted and normal hierarchy is interesting. Most notably for the unperturbed symmetric case, where we find the Lyapunov exponent in the normal hierarchy to be six times as large as the Lyapunov exponent in the inverted hierarchy. This is of course also comparing a stationary to a bipolar case which really are two different solutions, but even for the non-periodic cases there is a difference. The consequence is that perturbations in the normal hierarchy grow faster than similar perturbations in the inverted hierarchy.
If this pattern is also found in more realistic models, it might lead to significant differences between the normal and the inverted hierarchy in the early universe or a supernova. In these environments we find a decaying neutrino background potential with time or radius respectively, and this decay limits the region where perturbations can grow.
Therefore the growth rate will determine if the perturbation becomes large and changes the observable signatures.

Finally, while the results for this two beam model is of limited use when considering real physical systems, the Lyapunov analysis highlights that there is a lot of information about the stability of such a system to be found beyond a simple stationary linearization.



\section*{Acknowledgments}

We would like to thank Yvonne Wong for comments on the manuscript and Georg Raffelt for numerous valuable discussions and critical comments on earlier versions of the manuscript.

\appendix

\section{Lyapunov analysis}
\label{sec:lyapunov}


We will here define the Lyapunov exponents and covariant Lyapunov vectors and describe the most important information obtainable from them, but, first, we need to settle a few definitions regarding the linearization of differential equations.

Given a differential equation of the form
\begin{equation}
  \label{eq:1}
  \dot{\mathbf{y}}(t) = \mathbf{G}(\mathbf{y}(t)) ,
\end{equation}
we can linearize it and define the Jacobian $\mathbf{J}(t)$ as
\begin{equation}
  \label{eq:2}
  \dot{\mathbf{y}}(t+dt) \approx \mathbf{G}(\mathbf{y}(t)) + \mathbf{J}(t) d\mathbf{y} \equiv \mathbf{G}(\mathbf{y}(t)) + \frac{\partial \mathbf{G}(\mathbf{y})}{\partial \mathbf{y}} d\mathbf{y} .
\end{equation}
From this linearization, it can be shown that an infinitesimal perturbation $\mathbf{v}$ to the trajectory will evolve according to the differential equation
\begin{equation}
  \label{eq:vdiff}
  \dot{\mathbf{v}}(t) = \mathbf{J}(t)\mathbf{v}(t) .
\end{equation}
With this in mind, we define the propagator from $t_1$ to $t_2$, $\mathbf{M}(t_1,t_2)$, to be the linear operator which evolves any perturbation $\mathbf{v}$ from $t_1$ to $t_2$
\begin{equation}
  \label{eq:4}
  \mathbf{v}(t_2) = \mathbf{M}(t_1,t_2) \mathbf{v}(t_1) ,
\end{equation}
and with this definition, the propagator must obviously also be a solution to $\dot{\mathbf{M}}(t_1,t) = \mathbf{J}(t) \mathbf{M}(t_1,t)$.
Throughout Appendix~\ref{sec:lyapunov} and Appendix~\ref{sec:calc}, we use the convention that bold face lower case refers to vectors while bold face upper case refers to matrices. Since we do not refer to the polarization vectors from the main text at all, there should be no chance of confusion.

\subsection{Lyapunov exponents}
\label{subsec:lyaexp}

From the definitions above, it is possible to define a plethora of different characteristic numbers which are typically eigenvalues of some matrix. Of these, the most commonly used is the Lyapunov exponent. Given a perturbation $\mathbf{v}$, the associated Lyapunov exponent can be defined in several different ways, but the most intuitive one is~\cite{Ginelli2013, Wolf1986, Benettin1980}
\begin{equation}
  \label{eq:lya1}
  \lambda = \lim_{t' \rightarrow \infty} \frac{1}{t'} \ln \frac{|| \mathbf{v}(t') ||}{|| \mathbf{v}(t_0) ||} ,
\end{equation}
which was also introduced in \eref{eq:lambda}.
Choosing different perturbing vectors turns out to give rise to a spectrum of different Lyapunov exponents $\lambda_1 \ge \lambda_2 \ge \dots \ge \lambda_n$. 
A neat way to picture the different Lyapunov exponents is to consider a generalized box of dimension $k$ where the sides consist of vectors $\mathbf{v}_1$, $\mathbf{v}_2$, $\dots$, $\mathbf{v}_k$ corresponding to $\lambda_1$, $\lambda_2$, $\dots$, $\lambda_k$. Then the generalized volume will grow as $\exp(t\sum_{i=1}^k \lambda_i)$ if $\lambda_i \neq \lambda_j$ for $i \neq j$. In this way, the Lyapunov exponents describe how the dynamics deform a volume initially surrounding a given point on the trajectory. This view is explored further in Refs.~\cite{Wolf1986, Kuptsov2012}.

The Lyapunov exponents can also be defined as the logarithms of the eigenvalues of the matrix~\cite{Kuptsov2012, Wolfe2007}
\begin{equation}
  \label{eq:lya2}
  \mathbf{W}_+(t) = \lim_{t' \rightarrow \infty} \left[ \mathbf{M}(t,t')^T \mathbf{M}(t,t')\right]^{1/(2(t'-t))} .
\end{equation}
This limit exists for almost every $t$ under some quite weak assumptions according to Oseledets multiplicative theorem~\cite{Oseledets}, and the definition can also be related to \eref{eq:lya1} simply by using $\mathbf{v}(t') = \mathbf{M}(t_0,t') \mathbf{v}(t_0)$.

When the Lyapunov exponent is considered an eigenvalue, it is quite straight forward to define the multiplicity $m$ of a Lyapunov exponent to be the dimension of the associated eigenspace. 
$\mathbf{W_+}$ is symmetric and real, so the sum of all multiplicities must equal the dimension of the system, $n$. We will use the convention that summing over the Lyapunov exponents implicitly means repeating the degenerate ones $m$ times such that we always have $n$ Lyapunov exponents.

The Lyapunov exponents carry a lot of information about the system. For a bounded trajectory, $\lambda_1 > 0$ indicates that the system is chaotic since this indicates that the distance between initially nearby trajectories will diverge exponentially.
It can also be shown that the $\mathrm{trace}(\mathbf{W}_+) = \sum_{i=1}^n \lambda_i = 0$ if the system is conservative and thereby also invertible~\cite{Ginelli2013}. Similarly, the trace is negative if the system is dissipative~\cite{Wolf1986}. 

Another interesting connection exists between Lyapunov exponents, entropy, and information loss. 
Pesin has shown that the entropy of a system is the sum of positive Lyapunov exponents~\cite{Pesin1976,Ginelli2013,Eckmann1985} assuming that the system is ergodic.\footnote{A system is ergodic if the average is the same whether it is over time or over phase space, and it is a reasonable good assumption for the system considered here.} A more intuitive treatment is given by Wolf in Ref.~\cite{Wolf1986} where he argues that the Lyapunov exponents give the rate of information loss in bits per time unit if \eref{eq:lya1} is defined with $\log_2(x)$ instead of $\ln(x)$. We will, however, follow the literature and continue to use $\ln(x)$, and we will discuss how fast small perturbations will grow rather than considering information loss in bits per time unit.

Loss of information and growth of small perturbations really are two sides of the same coin. If we know a number to $k$ digits of precision, this corresponds to $k$ digits of information. On the other hand, it also constrains the largest perturbations to be in the order of $10^{-k}$.
Given a Lyapunov exponent $\lambda$, we lose $\log e^\lambda$ digits of information per time unit, and all our information would be lost in $k/\log e^\lambda$ time units. On the other hand, it means that the perturbation would grow to the order of one in $-\ln (10^{-k})/\lambda = k/\lambda\log e$ time units, so the two approaches give identical results. Given a certain perturbation, we also see that multiplying the Lyapunov exponent by a factor will shorten the time needed for the perturbation to grow with the same factor.

\subsection{Lyapunov vectors}
\label{subsec:lyavec}

The Lyapunov exponents tell us something about how unstable the system is, but it is also interesting to associate a direction with these instabilities. Such vectors are called Lyapunov vectors in general, and we will see a few different examples. An obvious choice as a Lyapunov vector could be the normalized eigenvectors of $\mathbf{W_+}(t)$~\cite{Goldhirsch1987}. These are called the forward singular vectors, and we will use $\mathbf{f}_{i}(t)$ for the forward singular vector corresponding to the $i$'th Lyapunov exponent.
The problem with the forward singular vectors is that they do not respect the dynamics of the system because $\mathbf{f}_{i}(t_2) \neq \mathbf{M}(t_1,t_2) \mathbf{f}_{i}(t_1)$~\cite{Kuptsov2012}. 
If we want our Lyapunov vectors to respect the dynamics of the system, we need to use the covariant Lyapunov vectors which also respect the time-reversed dynamics and generalize the stationary normal modes to arbitrary trajectories~\cite{Ginelli2007,Wolfe2007}.
In order to define the covariant Lyapunov vectors, we will first consider the Oseledets subspaces connected to the Lyapunov exponents~\cite{Wolfe2007,Kuptsov2012}. 
Let the subspace $O^+_i$ consist of all vectors in the tangent space where \eref{eq:lya1} gives a Lyapunov exponent smaller than or equal to $\lambda_i$. 
In terms of the forward singular vectors, the first set of Oseledets subspaces can be expressed as
\begin{equation}
  \label{eq:7}
  O^+_i = \textrm{span}\left\{\mathbf{f}_{j}(t) | j \ge i\right\} .
\end{equation}
For the time-reversed dynamics, we can define a matrix similar to $\mathbf{W}_+(t)$, only taking the limit of $t' \rightarrow -\infty$.
\begin{equation}
  \label{eq:lya3}
  \mathbf{W}_-(t) = \lim_{t' \rightarrow -\infty} \left[ \mathbf{M}(t,t')^T \mathbf{M}(t,t')\right]^{1/(2(t'-t))} .
\end{equation}
The eigenvalues of $\mathbf{W}_-(t)$ are $-\lambda_1, -\lambda_2, \dots, -\lambda_n$, and the corresponding normalized eigenvectors, $\mathbf{b}_{i}(t)$, are called the backward singular vectors. From these vectors, we can define the second set of Oseledets subspaces
\begin{equation}
  \label{eq:8}
  O^-_i(t) = \textrm{span}\left\{\mathbf{b}_{j}(t) | j \le i\right\} .
\end{equation}
These subspaces have the property that any vector in $O^-_i(t)$ will give a Lyapunov exponent smaller than or equal to $-\lambda_i$ when the system is evolved backwards in time.
The Oseledets subspaces can now be used to give a stringent definition of the covariant Lyapunov vectors.
For a vector to be in $O^+_i(t)$, it cannot grow faster than $\lambda_i$, so $\mathbf{M}(t_1,t_2) O^+_i(t_1) = O^+_i(t_2)$, and therefore $\g_i$, the covariant Lyapunov vector corresponding to $\lambda_i$, must be in this subspace. To fully respect the dynamics of the system, the covariant Lyapunov vector must also respect time reversal. For the reversed propagator we get that $\mathbf{M}^{-1}(t_1,t_2) O^-_i(t_2) = O^-_i(t_1)$, so $\g_{i}(t)$ must also be in the subspace $O^-_i(t)$. It turns out that
\begin{equation}
  \label{eq:9}
  \textrm{span}(\g_i(t)) = O^+_i(t) \bigcap O^-_i(t).
\end{equation}
Again it is worth noting that in the degenerate case all linear combinations of the individual covariant Lyapunov vectors corresponding to $\lambda_i$ should be considered, and each covariant Lyapunov vector can be chosen at will in $\textrm{span}(\g_j(t) | \lambda_j = \lambda_i)$ if only it is linearly independent from all of the others.

An important advantage of the covariant Lyapunov vectors is that they reduce to the normal modes of the system given the solution to the differential equation is stationary.
If they are found for a periodic trajectory, they similarly reduce to the so-called Floquet vectors, and in this way, the covariant Lyapunov vectors are simply the generalization of normal modes to arbitrary chaotic trajectories~\cite{Wolfe2007}.
The major disadvantage of covariant Lyapunov vectors is the lack of one to one correspondence between the non-zero components and diverging directions as we discussed at length in \sref{sec:lyalight}.

Finally, an application for the singular vectors and the so-called bred vectors (which we have not considered here)~\cite{Buizza2005}, and in the future possibly also covariant Lyapunov vectors, is to improve the efficiency of forecasting in multidimensional systems. This is possible since knowledge about the unstable directions can help to choose an optimal set of initial conditions for exploring the full space of solutions.



\section{Numerical calculation of Lyapunov exponents and Covariant Lyapunov vectors}
\label{sec:calc}

The first numerical algorithm to calculate the full spectrum of Lyapunov exponents was proposed by Benettin et al.~\cite{Benettin1980, Benettin}. The leading exponent can easily be calculated by simply choosing a random vector in the tangent space and evolving it according to \eref{eq:vdiff}. The only complication is that the components of the vector can exceed the value admissible for a float or a double. The ease of finding the leading Lyapunov exponent is also the curse for finding any of the other exponents. If the equations can be inverted, it is possible to find the smallest Lyapunov exponent from the inverted dynamics, but everything in between needs a trick.

The trick is to orthogonalize the vectors before they collapse into the most unstable direction.  This can be done using a standard $\mathbf{QR}$-decomposition where the unitary $\mathbf{Q}$-matrix contains the orthonormalized vectors, and the diagonal of the upper triangular $\mathbf{R}$-matrix contains the lengths of the orthogonalized vectors before they are normalized. These lengths are exactly what is needed in order to compute the Lyapunov exponents. For a trajectory divided in $k$ sections, the Lyapunov exponents are given by
\begin{equation}
  \label{eq:10}
  \lambda_i \approx \frac{1}{t} \ln\left(\prod_{j=1}^k r_{jj}\right) = \frac{1}{t} \sum_{j=1}^k \ln r_{jj} .
\end{equation}
This is a finite time version of \eref{eq:lya1} where $||v(t_0)||=1$ since we orthonormalize the vectors in each step.

\begin{figure}[tbp]
\center
\includegraphics[width=\columnwidth]{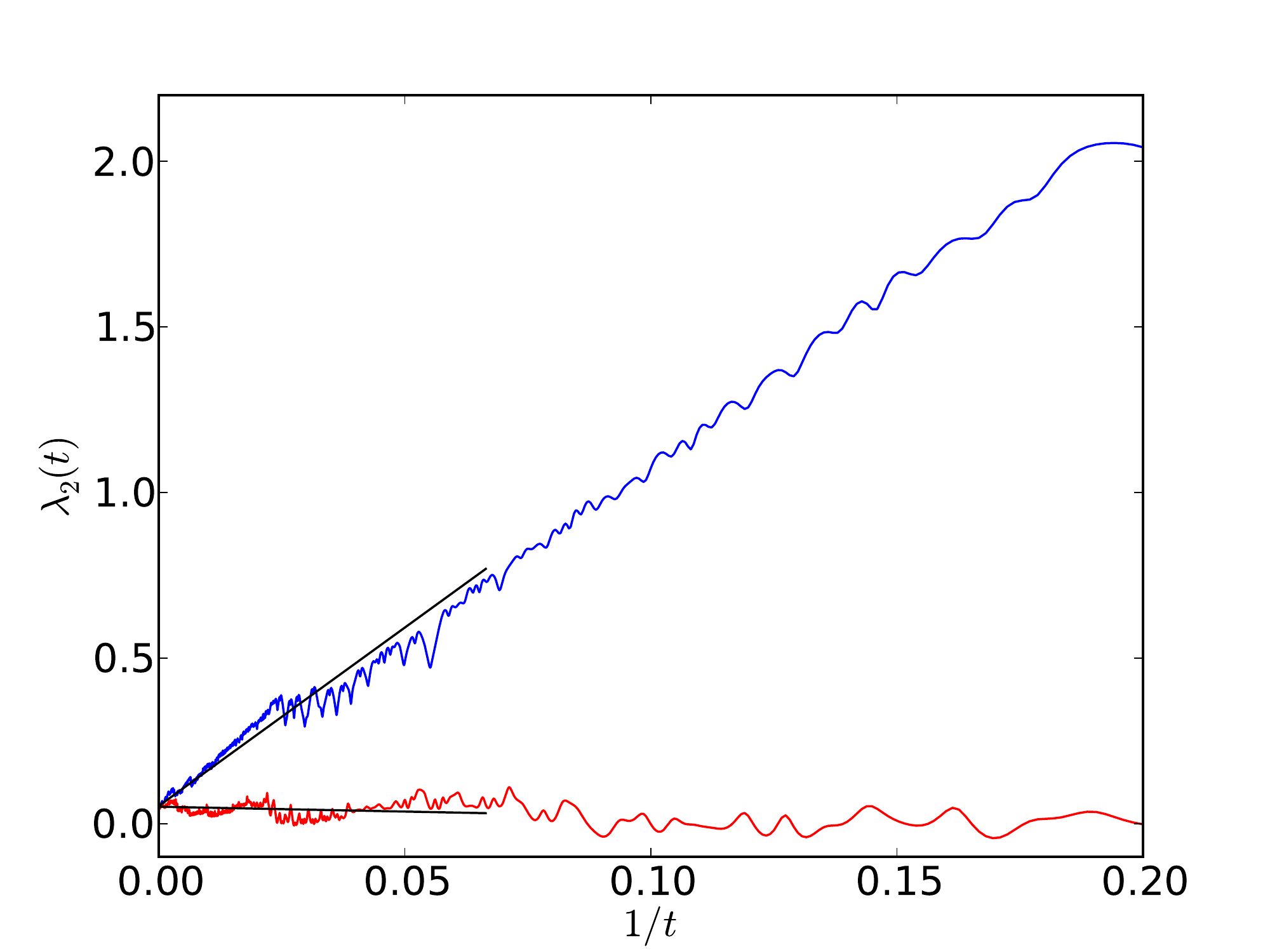}
\caption{The asymptotic value of $\lambda_2$ is found at $1/t = 0$. The blue line shows the Lyapunov exponents obtained from the forward calculation for different time intervals while the red line shows the Lyapunov exponents obtained from the backward calculation. The black lines are fits where the first 300 time steps have been excluded. This plot is for the inverted hierarchy with the modification $\delta S_z$ and $S_{1x} = - S_{2x}$, and with the procedure for calculating the error described in the text, we find $\lambda_2 = 0.045\pm0.003$.}
\label{fig:lya_conv}
\end{figure}

\begin{figure}[tbp]
\center
\includegraphics[width=\columnwidth]{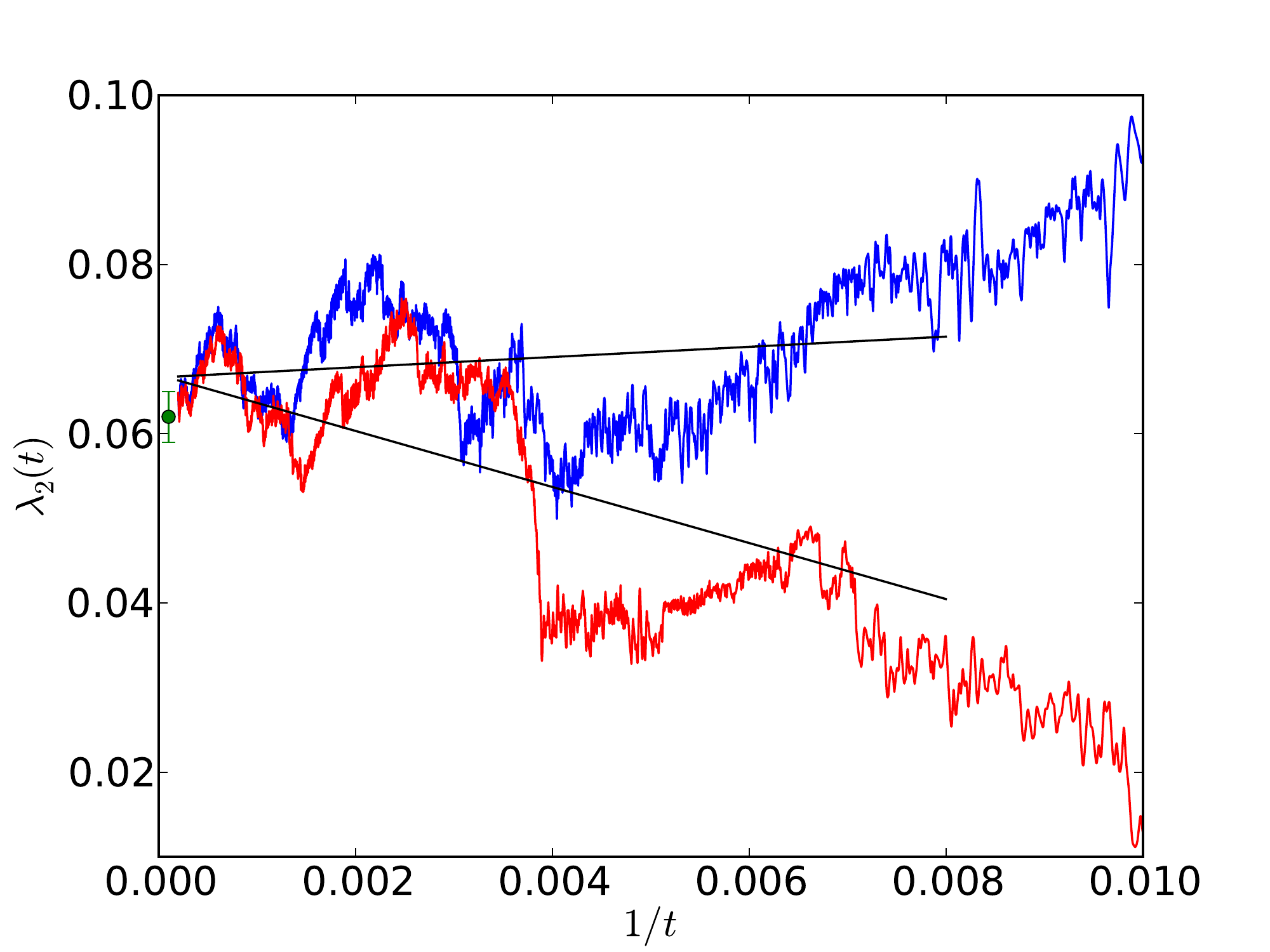}
\caption{The asymptotic value of $\lambda_2$ is found at $1/t = 0$. The blue line shows the Lyapunov exponents obtained from the forward calculation for different time intervals while the red line shows the Lyapunov exponents obtained from the backward calculation. The black lines are fits where the first 2500 time steps have been excluded. This plot is for the inverted hierarchy with the modification $\delta S_y$ and $S_{1x} = -S_{2x}$, and with the procedure for calculating the error described in the text, we find $\lambda_2 = 0.062\pm0.003$ marked by a green point in the figure.}
\label{fig:ih_sy_m_le_2}
\end{figure}

It turns out that in many cases, the convergence of $\lambda_i$ is quite slow. In order to improve this, Goldhirsch et al.~\cite{Goldhirsch1987} showed that the error depends on time as $1/t$. This means that plotting $\lambda_i(t)$ versus $1/t$ should yield a straight line where the asymptotic value for $\lambda_i$ is found at $1/t=0$. An example of this is seen in \fref{fig:lya_conv} where we show two sets of calculated Lyapunov exponents.
As it can be seen in the Figure, the estimated asymptotic value will depend somewhat on the amount of data that is used when fitting the straight line. This is even more clear in \fref{fig:ih_sy_m_le_2} where we show a smaller segment of the time axis. We have used this difference as a measure of the error in the computed Lyapunov exponents. The computed set of values contain 101000 time steps, and we have constructed a sample of different estimates of $\lambda_i$ by calculating $\lambda_i$ 500 times using the last 100900 to 51000 time steps. This is done for both the sets of values shown in \fref{fig:lya_conv}, and, finally, we calculated the mean and the standard deviation of this full sample to get $\lambda_i \pm \epsilon$. For \fref{fig:lya_conv} this gives $0.0455 \pm 0.003$, and for \fref{fig:ih_sy_m_le_2} it gives $0.062\pm0.003$. Note that the different values in our samples are not independent, and thus the error we calculate should not be interpreted as a stringent standard deviation but rather as an indication of our level of precision. The lack of independence is partly due to the origin of the two different sets of values in \fref{fig:lya_conv}, but before we describe that, we need to understand how the covariant Lyapunov vectors are calculated.

The algorithm used to find the covariant Lyapunov vectors is a slight modification of the one presented by Kuptsov and Parlitz~\cite{Kuptsov2012}. The idea builds mainly on what was proposed by Wolfe and Samelson~\cite{Wolfe2007} while Ginelli et al.~\cite{Ginelli2007} has an alternative approach.

The method we use to compute covariant Lyapunov vectors is to find the forward and backward singular vectors first and then calculate the covariant Lyapunov vectors.
To find the backward singular vectors, we can consider an arbitrary vector $\mathbf{v}$ in the tangent space. Almost any such vector will grow with the average rate $\lambda_1$ giving $|\mathbf{v}(t_2)| = |\mathbf{v}(t_1)|\exp(\lambda_1(t_2-t_1))$. 
Using the time-reversed dynamics on $\mathbf{v}(t_2)$, we find
\begin{equation}
  \label{eq:3}
\mathbf{M}^{-1}(t_1,t_2) \mathbf{v}(t_2) = \mathbf{M}^{-1}(t_1,t_2) \mathbf{M}(t_1,t_2) \mathbf{v}(t_1) = \mathbf{v}(t_1) ,
\end{equation}
so $\mathbf{v}(t_2)$ contracts at a rate $-\lambda_1$ under the backwards dynamics and therefore approaches $\mathbf{b}_{1}(t_2)$ as we use longer time intervals. We assume that all Lyapunov exponents are non-degenerate, but the generalization to the degenerate case is straight forward.
Considering an area spanned by $\mathbf{v}$ and another arbitrary vector $\mathbf{u}$, again for almost any choice of $\mathbf{u}$, it will grow with the rate $\lambda_1+\lambda_2$. This means that the component of $\mathbf{u}$ orthogonal to $\mathbf{b}_1$ must approach $\mathbf{b}_2$ by an argument similar to the one above. This process can be repeated, and we can find all the backward singular vectors in this way by induction.
Going back to our method for calculating the Lyapunov exponents, it should be clear that a byproduct of the calculation is that the backward singular vectors end up as the columns in the $\mathbf{Q}$-matrix of the $\mathbf{QR}$-decomposition.

With a more involved argument, it is also possible to show that the forward singular vectors can be obtained by using $\mathbf{M}^T(t_1,t_2)$ and going backwards in time~\cite{Kuptsov2012}.

Having obtained the forward and backward singular vectors, we need a method to find the covariant Lyapunov vectors. Since the covariant Lyapunov vectors must respect both forward and backward dynamics, it must be possible to write the matrix $\mathbf{\Gamma}(t) = [\g_{1} ... \g_{n}]$ as
\begin{equation}
  \label{eq:Gamma}
  \mathbf{\Gamma}(t) = \mathbf{B}(t) \mathbf{A}^-(t) = \mathbf{F}(t) \mathbf{A}^+(t) ,
\end{equation}
where $\mathbf{B}(t)$ and $\mathbf{F}(t)$ are the matrices of backward and forward singular vectors respectively. As the $i$'th covariant Lyapunov vector must grow with only $-\lambda_i$ in the backwards dynamics, it can only have components from $\mathbf{b}_{j}$ with $\lambda_j \ge \lambda_i$. This means that $\mathbf{A}^-(t)$ can be chosen to be upper diagonal. Similarly $\mathbf{A}^+(t)$ can be chosen to be lower diagonal.
Multiplying by $\mathbf{F}^T(t)$, we get the equation
\begin{gather}
  \label{eq:ApAm}
  \mathbf{F}^T(t) \mathbf{B}(t) \mathbf{A}^-(t) = \mathbf{A}^+(t) \Leftrightarrow\\
  \mathbf{F}^T(t) \mathbf{B}(t) = \mathbf{A}^+(t) (\mathbf{A}^-(t))^{-1} ,
\end{gather}
which is a LU-factorization of $\mathbf{F}^T(t) \mathbf{B}(t)$. 
To find $\mathbf{A}^-(t)$, we can restrict our attention to the upper left $j$ times $j$ submatrix of $\mathbf{F}^T(t) \mathbf{B}(t)$ in \eref{eq:ApAm} and focusing on the $j$'th column of $\mathbf{A}^\pm(t)$. This gives an equation of the form
\begin{equation}
  \label{eq:Aminus}
  \begin{pmatrix}
  x_{11} & x_{12} & \dots & x_{1j}\\ 
  x_{21} & x_{22} & \dots & x_{2j}\\ 
  \dots & \dots & \dots & \dots\\ 
  x_{(j-1)1} & x_{(j-1)2} & \dots & x_{(j-1)j}\\
  x_{j1} & x_{j2} & \dots & x_{jj} 
  \end{pmatrix}
  \begin{pmatrix}
  a^-_{1j}\\a^-_{2j}\\\dots\\a^-_{(j-1)j}\\a^-_{jj}
  \end{pmatrix}
  =
  \begin{pmatrix}
  0\\0\\\dots\\0\\a^+_{jj}
  \end{pmatrix}
  .
\end{equation}
Since the LU-factorization is unique only up to the diagonal of one of the matrices, we can eliminate the $j$'th row in the above matrix equation resulting in a homogeneous system. Solving this to find $\mathbf{A}^-(t)$, we can find $\mathbf{\Gamma}(t)$ from \eref{eq:Gamma}.

We now have all the tools to find the covariant Lyapunov vectors, but let us go back and get the full overview of the numerical algorithm. 
To control the divergence of the singular vectors, we solve the differential equations for short time steps $dt$ (we use $dt=0.05$) and use the result as the initial conditions for the next step. The calculation goes through four different phases:

\begin{enumerate}

\item 
\label{item:init}
A random unitary matrix initialize $\mathbf{B}$. The trajectory and propagator is found for each time step, and $\mathbf{B}$ is evolved by using the propagator and finding the $\mathbf{QR}$-factorization of the result. The $\mathbf{R}$-matrix diagonal is saved. (We used 1000 steps)

\item 
\label{item:forward}
The trajectory and propagator is found for each time step, and $\mathbf{B}$ is evolved as before. The $\mathbf{R}$-matrix diagonal, the trajectory, and $\mathbf{B}$ are saved. (We used 100000 steps)

\item
\label{item:binit}
The trajectory is found and saved for each time step as preparation. A random unitary matrix initialize $\mathbf{F}$. Starting with the last time step, the trajectory and propagator is found from the previous time step and evolved to the current one. $\mathbf{F}$ is then evolved backwards in time by using $\mathbf{M}^T$ on $\mathbf{F}$ and finding the $\mathbf{QR}$-factorization of the result. The $\mathbf{R}$-matrix diagonal is saved. (We used 1000 steps)

\item
\label{item:backward}
Starting with the last time step from phase~\ref{item:forward}, $\mathbf{F}$ is evolved backwards in time as described in phase~\ref{item:binit}. Knowing both $\mathbf{B}$ and $\mathbf{F}$, \eref{eq:Gamma} and \eref{eq:Aminus} gives the covariant Lyapunov vectors. The $\mathbf{R}$-matrix diagonal and the covariant Lyapunov exponents are saved. (We used 100000 steps as in phase~\ref{item:forward})

\end{enumerate}

In this way, we find the covariant Lyapunov vector, and it is possible to save the forward singular vectors if we are interested in those. Furthermore, we obtain two sets of data from which we can estimate the Lyapunov exponents. The caveat here is that these two sets of data are not independent as they originate in the same trajectory. In order to eliminate this dependency, we would need to skip one of the data sets, and thereby we would loose some of our precision.



\bibliography{twobeam}

\end{document}